\documentstyle[epsf]{elsart}
\journal{Physica D}
\date{14 February 1998}
\begin{document}
\begin{frontmatter}
\title{%
On the Tongue-Like Bifurcation Structures 
of the Mean-Field Dynamics 
in a Network of Chaotic Elements}
\author[shibata]{Tatsuo Shibata},
\author[kaneko]{Kunihiko Kaneko}
\address{%
Department of Pure and Applied Sciences, 
University of Tokyo,
Komaba, Meguro-ku, Tokyo 153, Japan}
\thanks[shibata]{shibata@complex.c.u-tokyo.ac.jp}
\thanks[kaneko]{kaneko@cyber.c.u-tokyo.ac.jp}

\begin{abstract}
Collective behavior is studied in globally coupled maps. Several
coherent motions exist, even in fully desynchronized state.  To
characterize the collective behavior, we introduce scaling
transformation of parameter, and detect the tongue-like structure of
collective motions in parameter space.  Such collective motion is
supported by the separation of time scale, given by the
self-consistent relationship between the collective motion and chaotic
dynamics of each element.  It is shown that the change of collective
motion is related with the window structure of a single
one-dimensional map. Formation and collapse of regular collective
motion are understood as the internal bifurcation structure.
Coexistence of multiple attractors with different collective behaviors
is also found in fully desynchronized state.

05.45+b,05.90+m,87.10+e
\end{abstract}
\end{frontmatter}


\section{Introduction}

Whereas the research of low dimensional chaos provided us with
important notion of unpredictability in deterministic systems, it was
soon realized that many natural systems are much more complicated than
the low dimensional chaos.  One of the important features in such
system is high dimensionality.  Although there remains deterministic
aspects in the high dimensional chaos, the present nonlinear dynamics
tools are not sufficient to distinguish it clearly from noise.  Hence,
the study of high-dimensional chaos is important both from theoretical
and practical points of views.

Globally coupled dynamical systems, which consist of many dynamical
elements interacting all-to-all, are a good example to develop notions
in high dimensional systems.  Such a class of dynamical systems is
seen in physical, chemical and biological systems.  In physics,
coupled Josephson junction array\cite{Watanabe} is a coupled nonlinear
oscillator circuit with a global feedback. In nonlinear optics with
multi-mode excitation\cite{opt.} many modes are often coupled globally
through energy currency.  In bioscience and medical science,
neural\cite{Aertsen}, cellular\cite{Ko}, and vital\cite{Godin}
organizations are considered as a network of active elements which are
known to exhibit complex chaotic behaviors.  Several examples in
ecological and economic systems are also considered as a network of
active agents.  Globally coupled dynamical systems is the simplest
model among these complex network of active elements.

So far, study of globally coupled dynamical systems has revealed novel
concepts\cite{Kaneko1990a} such as clustering, chaotic itinerancy, and
partial ordering.  In particular, study of collective dynamics has
gathered much attention\cite{Kaneko1990b,Kaneko1992,Perez1992,%
Chate1992,Perez1993,Pikovsky1994,Just1995a,Kaneko1995,Ershov1995,%
Morita1996,Chate1996,Shibata1997,Ershov1997,Nakagawa1998,Chawanya}.
When the interactions between elements are small enough, each element
oscillates independently without synchronization between
them. Therefore the degrees of freedom of the system are effectively
proportional to the system size. If each element has chaotic dynamics,
the system may be thought as high dimensional chaotic state.  Even in
such a case, a macroscopic variable show some kind of complicated
dynamics rather than noise, ranging from low-dimensional torus to
high-dimensional chaos\cite{Shibata1997,Pikovsky1994}.  This may imply
that any weak interaction between active elements necessarily brings
some sort of correlation among them.

The purpose of the present paper is to study the nature of such
collective motion adopting a globally coupled map\cite{Kaneko1990a},
and present a mechanism for the origin of such collective dynamics.
With the change of the control parameters, collective dynamics shows
some sort of bifurcation. We present how elements are organized to
show the bifurcation structure in the collective dynamics.

In Section \ref{sec:model}, globally coupled logistic map is
introduced and its characteristic phenomena are presented as a brief
review.
In Section \ref{sec:phenomena}, an overview of different kind of
collective dynamics in the desynchronized state of globally coupled
logistic maps is presented. In macroscopic dynamics, lower dimensional
motion and much longer time scale than that of microscopic dynamics
are observed.

Our interest is focused on the thermodynamic limit of such collective
behavior.  In Section \ref{sec:thermo}, the time scale and the
amplitude of collective motion are studied in the limit of large
system size.  In Section \ref{sec:PhaseDia}, global phase diagram in
the parameter space is presented. While the phase diagram shows a
complicated structure, tongue-like bifurcation structures are
clarified by introducing a scaled nonlinearity parameter. Collective
dynamics with a larger amplitude exists in each tongue structure that
corresponds to a periodic window in the single logistic map.  The
elements are accumulated to few bands corresponding to the window for
a small coupling.  Since windows exist in any neighborhood in the
parameter space, the clarification of the collective dynamics with
such bands is necessary to understand the collective dynamics in
general.  Thus we focus on such tongue structures in Section
\ref{sec:selfconsistent}, to reveal a mechanism of collective dynamics,
where internal bifurcation of elements plays a key role.
In Section \ref{sec:bif}, bifurcation of tongue structure is studied
in connection with the internal bifurcation.
Even within the same tongue structure, we can observe different types
of collective motion. The growth of tongue structure with the
coupling strength is also discussed.
In Section \ref{sec:hysteresis}, hysteresis and multiple attractor
phenomena of the collective motion are reported.
This paper concludes in Section \ref{sec:end} with summary and
discussion.

\section{A Simple Network Model of Chaotic Elements on Globally
Coupled Map}
\label{sec:model}

In the present paper the following Globally Coupled Map(GCM) is
studied,
\begin{equation}
x_{n+1}(i)=(1-\epsilon)f(x_{n}(i))+\frac{\epsilon}{N}\sum_{j=1}^{N}f(x_n(j)),
\,\,\, (i=1, 2, 3, \cdots N), 
\end{equation}
where $x_{n}(i)$ is the variable of the $i$th element at discrete time
step $n$, and $f(x)$ is the internal dynamics of each element.  For
the internal dynamics we choose the logistic map
\begin{equation} 
f(x) = 1-a x^2,
\end{equation}
where $a$ is the nonlinearity parameter.  The logistic map has been
studied in detail as a typical of dissipative chaos.  The parameter
$\epsilon$ gives a coupling strength between elements.  The total
number of elements denoted as $N$.  The nonlinearity parameter $a$,
the coupling strength $\epsilon$, and the system size $N$ are the
control parameters of the GCM.

The GCM can be considered to be a mean-field extension of coupled map
lattice(CML)\cite{CML}, in which elements are located at discrete
spatial coordinates and interact with neighbors. GCM can be also
considered as a CML in which the spatial dimension goes to infinity.

In the GCM model, two opposite tendencies coexist: all-to-all coupling
tends to synchronize elements, while chaotic instability in each
element tends to desynchronize them.  Depending on the balance between
the two tendencies, a rich variety of phenomena has been found
\cite{Kaneko1990a}.  When the coupling strength is strong enough, all
elements are synchronized each other and the dynamics is nothing more
than the single logistic map as is called {\bf coherent phase}.  As
the coupling strength is smaller or the nonlinearity larger, elements
split into some groups, in each of which they are synchronized each
other.  This regime is called {\bf ordered phase}, while the phenomena
are called {\bf clustering}.  The clustering is common characteristics
in globally coupled systems, including globally coupled oscillator
systems\cite{Okuda,Nakagawa1993}.

In the region (called {\bf partially ordered phase}) where the two
opposite tendencies are somewhat balanced, some part of the elements
makes a few clusters, while the rest elements do not form clusters and
their oscillations are desynchronized.  In the phase space, there are
a lot of ``attractor ruins'' with lower dimensionality, at which the
trajectory is attracted and stays over some duration, but then the
trajectory goes out from them into much higher dimensional phase
space, till they are again attracted to another attractor ruin.  In
this phenomenon, called {\bf chaotic itinerancy}, effective degrees of
freedom changes with time\cite{Kaneko1990a,Kaneko1997,CI}.

If the coupling strength $\epsilon$ is small enough, desynchronizing
tendency is so dominant that elements are mutually
desynchronized\cite{Kaneko1990b,Kaneko1992}(called {\bf desynchronized
phase}).  In this case, in general, all Lyapunov exponents are
positive\footnote{Number of negative Lyapunov exponents may be related
to the number of synchronization\cite{Kuramoto}.} and the degrees of
freedom in this system are proportional to the system size $N$.  In
the desynchronized state, the mean-field is not stationary, where a kind
of collective dynamics is observed. In this paper we focus on the
macroscopic state on this desynchronized state.  To see the
macroscopic state, the dynamics and statistics of the mean-field
\begin{equation}
h_{n}=\frac{1}{N}\sum_{j=1}^{N}f(x_{n}(j)),
\end{equation}
are studied as an order parameter.  In the next section, we will show
some phenomena of macroscopic dynamics in the desynchronized state.

\section{Phenomenology of Collective Motion in Desynchronized State}
\label{sec:phenomena}

\begin{figure}
{\epsfxsize\textwidth\epsfbox{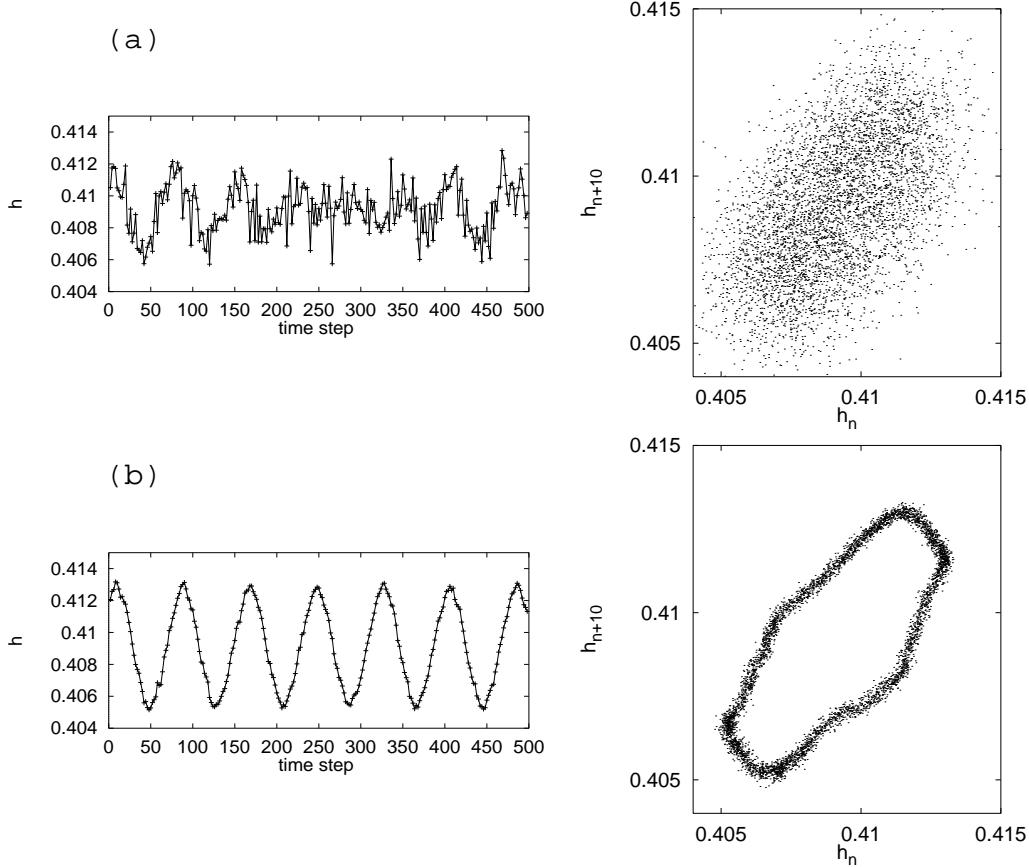}}
\caption{Time series and return map. Time series are plotted at every 2 
steps after transients are discarded. The parameters are
(a)$a=1.5449205$, $\epsilon=0.0005$, $N=10^5$, and
(b)$a=1.5449205$,$\epsilon=0.0005$, $N=10^7$. Corresponding return
maps $(h_n,h_{n+10})$ are plotted over 50000 steps after transients
are discarded.}
\label{fig:rm.p2}
\end{figure}

\begin{figure}
{\epsfxsize.75\textwidth\epsfbox{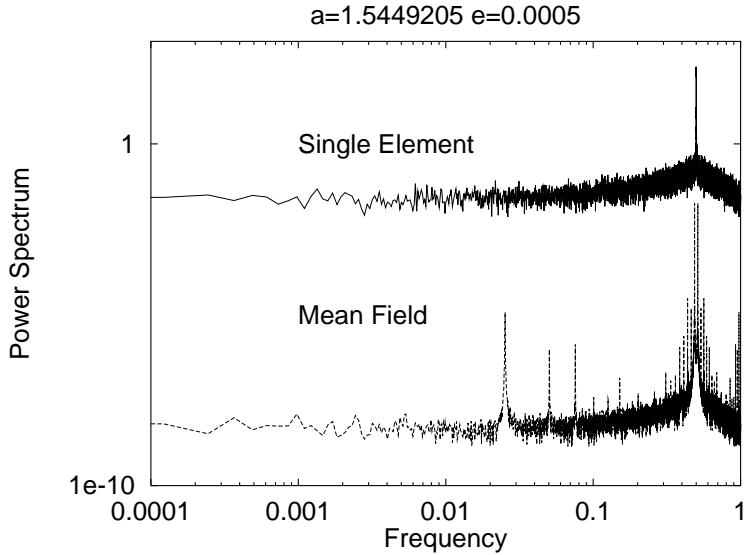}}
\caption{Power spectrum of time series of a single element $x_n(i)$(upper)
and the mean field $h_n$(lower), $a=1.5449205$, $\epsilon=0.0005$,
$N=10^7$, While the spectrum for a single element has a peak at the
frequency 0.5, the dynamics is more irregular than the mean
field. The slow dynamics of the mean field is shown at the frequency
0.025269.}
\label{fig:power}
\end{figure}

\begin{figure}
{\epsfxsize\textwidth\epsfbox{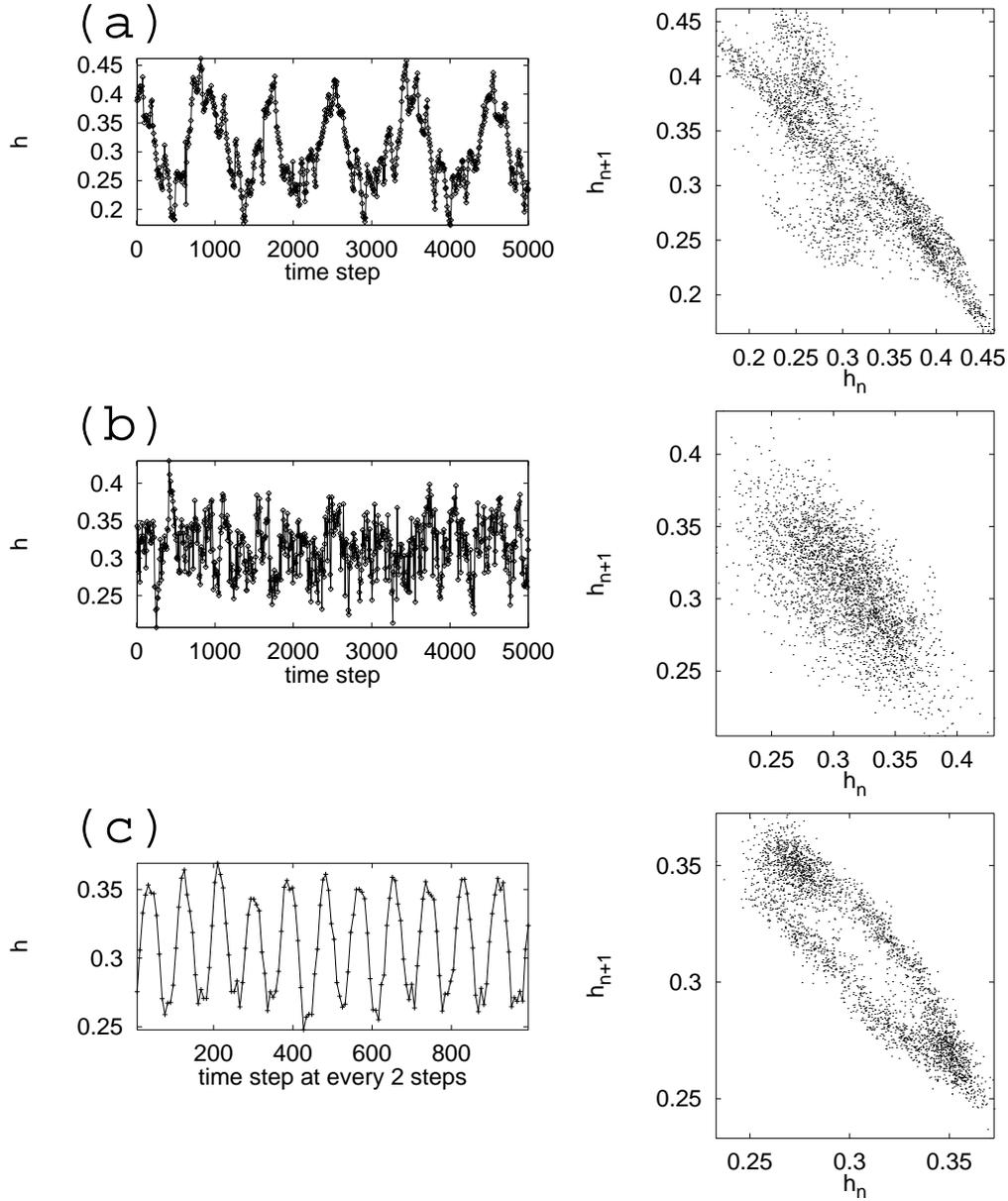}}
\caption{Time series and return map. Time series are plotted at every 7 
steps. The parameters are $a=1.69620$, $\epsilon=0.008$, $N=10^7$(a),
$a=1.69755$, $\epsilon=0.008$, $N=10^7$(b), $a=1.69844$,
$\epsilon=0.008$, $N=10^7$(c). See also Fig.\protect\ref{fig:power2}}
\label{fig:rm.p7}
\end{figure}

\begin{figure}
{\epsfysize.9\textheight\epsfbox{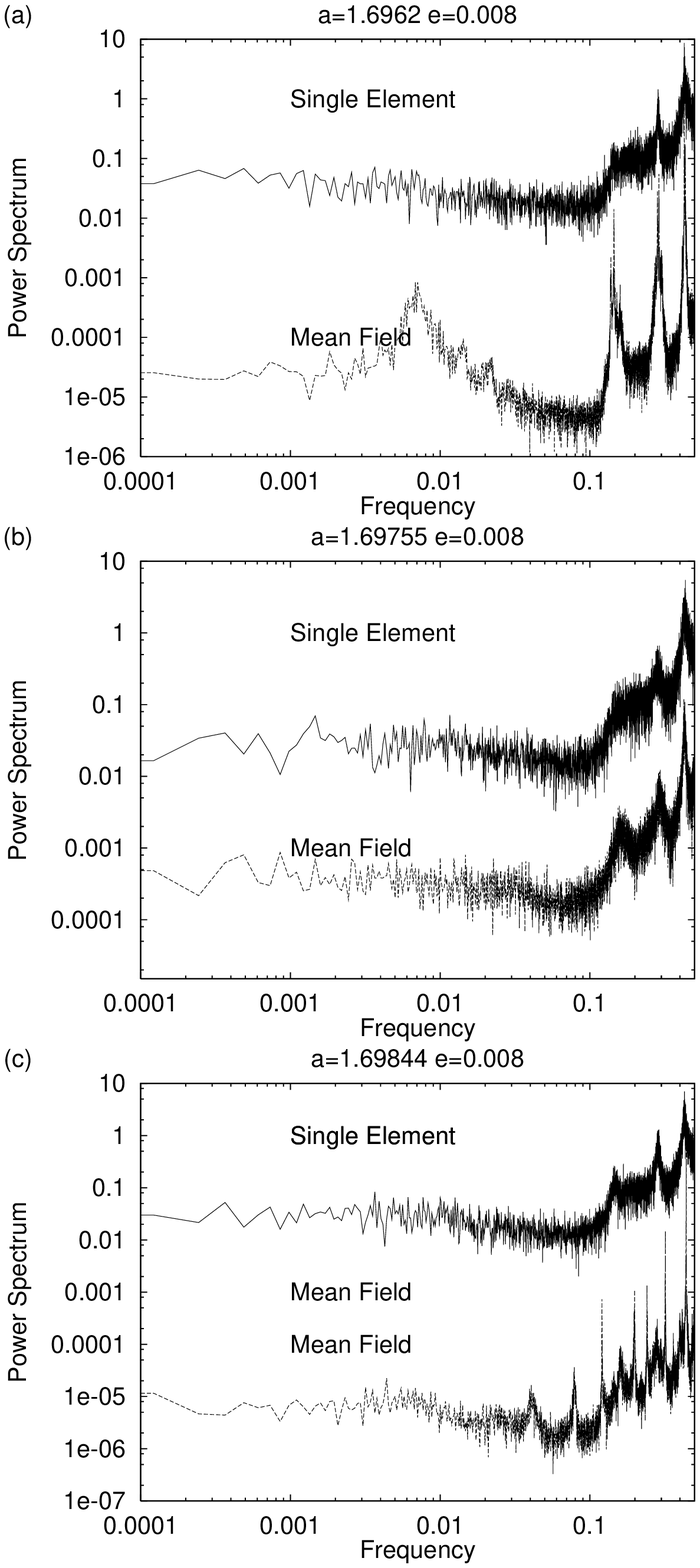}}
\caption{Power spectrum of time series of a single element
$x_n(i)$(upper) and the mean field $h_n$(lower), corresponding to
Fig.\protect\ref{fig:rm.p7}.}
\label{fig:power2}
\end{figure}

Since in the desynchronized state there is no mutual synchronization
in elements, one might imagine that the mean-field would be
effectively the same as noise and therefore the mean-field goes to a
constant with the increase of $N$.  One might consider that such high
dimensional dynamics can not be distinguishable from noise.

Indeed this is not the case.  One of the authors has found that the
mean field dynamics is different from noise, and studied its nature as
``hidden coherence''\cite{Kaneko1990b,Kaneko1992}. A simple solution
to such collective dynamics is the possibility that it is represented
by low-dimensional dynamics in the thermodynamic limit ($N \rightarrow
\infty$), even though each element is chaotic and mutually
desynchronized.  Indeed such examples have recently been found in
short-ranged coupled map lattice and cellular
automata\cite{Chate1992}, globally coupled
oscillators\cite{Nakagawa1994,Chabanol}, globally coupled tent
map\cite{Pikovsky1994,Morita1996,Nakagawa1998,Chawanya}, and globally
coupled logistic map with heterogeneous elements\cite{Shibata1997}.

In the present case, the collective dynamics is not given by such low
dimensional dynamics\cite{Kaneko1992}, although it has some structure
distinguishable from noise. Let us give a few sets of examples of the
mean-field dynamics.

Fig.\ref{fig:rm.p2}(a) shows the time series of the mean-field as a
function of time step $n$ at every 2 steps, and the corresponding
return map of the mean field (for $a=1.5449205, \epsilon=0.0005,
N=10^5$).  The coupling strength is too small to synchronize any two
elements.  The trajectory of the mean-field has some fluctuation due
to the finite system size.  With the increase of the system size $N$
(parameters are $a=1.5449205, \epsilon=0.0005, N=10^7$), however, the
trajectory shows some coherent motion as is shown in
Fig.\ref{fig:rm.p2}(b).  The trajectory is rather close to
quasiperiodic motion, although the points are scattered around the
``torus'' motion.  In Fig.\ref{fig:power}, power spectra for the
time series of an element and the mean-field are overwritten. The
mean-field dynamics has a much longer time scale than that of an
element.

Note that the width around the closed curve remains finite with the
further increase of $N$.  The collective dynamics is not on a
two-dimensional torus, and indeed is not represented by
low-dimensional dynamics as will be demonstrated in the next section.
On the other hand, since the mean-field dynamics does not approach a
point with the increase of $N$, it is also different from noise.
Hence the collective motion has some structure, although it is
high-dimensional.

Another set of examples is given in Fig.\ref{fig:rm.p7}, which are the
time series plotted at every 7 steps and the first return maps.  In
Fig.\ref{fig:rm.p7}(a) (parameters are $a=1.69620,
\epsilon=0.008, N=10^7$), quasi-periodic-like motion is not detected
in the mean field dynamics, but some structure exists in the return
map, while in the time series, characteristic time scale seems to
exist.

By slight increase of $a$ (i.e., with the parameters $a=1.69755,
\epsilon=0.008, N=10^7$), the dynamics of the mean field is changed as
in Fig.\ref{fig:rm.p7}(b).  In this case, the return map does not show
a clear structure, and the variation of the mean field remains at the
same magnitude with the further increase of $N$. 

With much slighter increase of $a$ (parameters are $a=1.69844$,
$\epsilon=0.008$, $N=10^7$), the mean field comes to oscillate more
regularly, whereas the motion is scattered around torus
motion(Fig.\ref{fig:rm.p7}).

Note that the choice of every 2 or 7 step in the above plots is not
arbitrary but there is a reason for it, as will be clear in the
following sections.  Our goal in this paper is to give a consistent
explanation for the above collective motion, and answer the remaining
questions in the collective dynamics: When the system size $N$ goes to
infinity, i.e., in the thermodynamic limit, how is the macroscopic
dynamics characterized?  How does the remnant order in
high-dimensional collective dynamics emerge out of the complete
desynchronized elements?  How is a longer-time scale in the collective
dynamics formed?  How does the collective dynamics depend on the
parameters $a$ and $\epsilon$, and what kind of bifurcation structure
is expected, and how is it explained in terms of dynamical systems
theory?

\section{Thermodynamic Limit of Collective Motions}
\label{sec:thermo}

\subsection{Amplitude of Collective Motion}

\begin{figure}
{\epsfxsize\textwidth\epsfbox{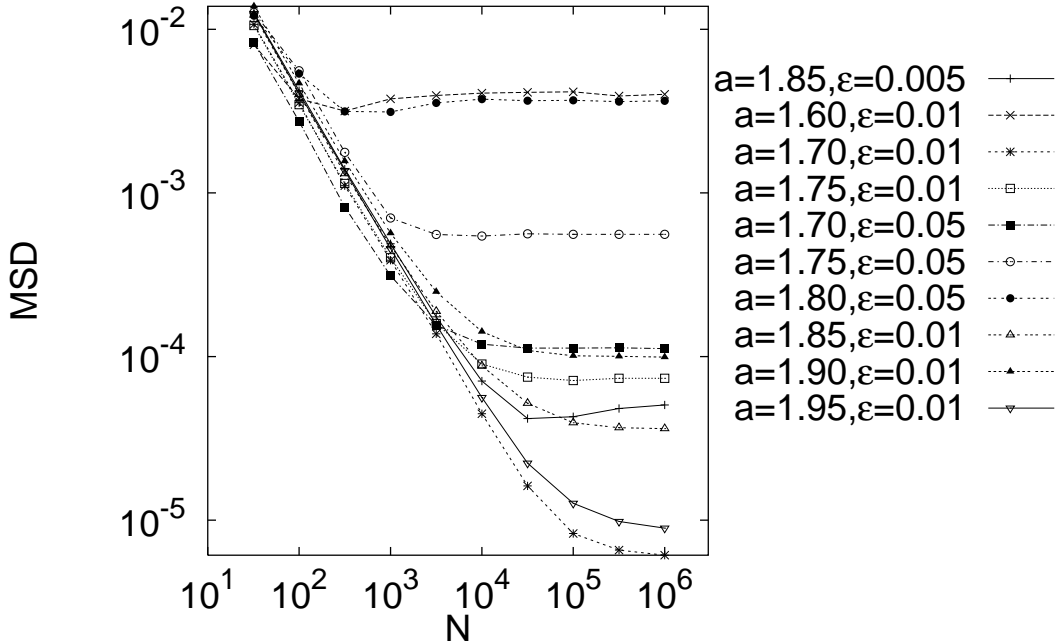}}
\caption{Mean square deviation (MSD) of the mean field distribution
are plotted as a function of the system size $N$.}
\label{fig:MSDvsN}
\end{figure}

In the previous section, we have mentioned that the mean field
dynamics plotted in the return map shows some structure, ranging from
lower-dimensional structure, such as torus, to higher-dimensional
stochastic structure. First of all, to characterize the mean
field dynamics, we measure the mean square deviation(MSD) of the mean
field distribution,
\begin{equation}
\langle(\delta h)^2\rangle =  \langle h^2 \rangle -\langle h \rangle^2,
\end{equation}
as a measure of the amplitude of the mean field dynamics. The bracket
$\langle\cdot\rangle$ denotes the temporal average. Since the motion
of the mean field is not on a torus, it is not always possible to
define the amplitude of oscillation by the radius of the torus
pattern.  Even though such collective oscillation is hardly detected,
the above MSD works as a measure for the amplitude, and also is useful
to measure the variation around the fixed point\footnote{Since there
is no synchronization each other, the MSD also provides a simple tool
to see whether the population obeys the law of large numbers.}.

In Fig.\ref{fig:MSDvsN} the MSD of the mean-field is plotted with the
system size $N$.  In some case, the MSD decreases up to a certain
constant with the increase of $N$, but remains constant with the
further increase.  In other cases, the increase of MSD is seen at some
range of size, but then approaches a certain constant.  These show the
distinction of the mean field dynamics from pure noise and suggest some
coherence between elements.

\subsection{Degrees of Freedom of Collective Motion}

\begin{figure}
{\epsfxsize\textwidth\epsfbox{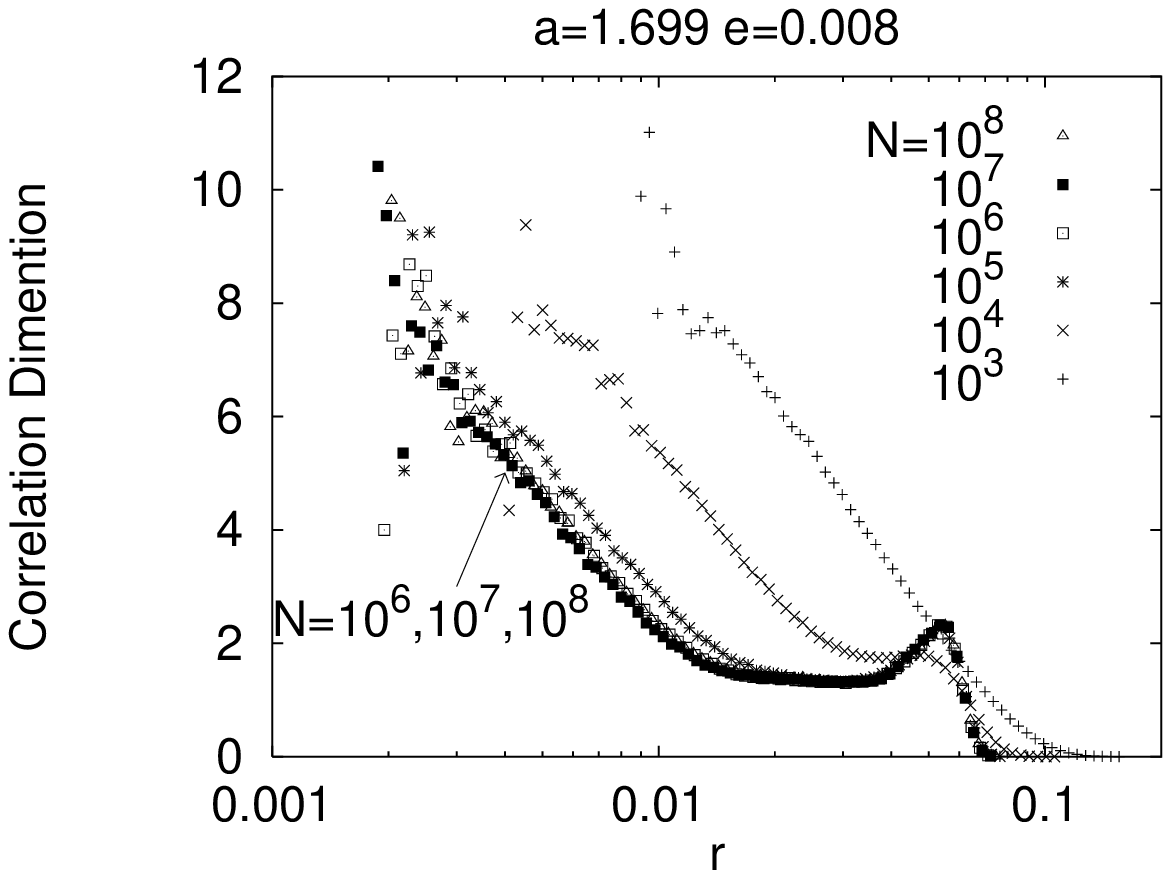}}
{\epsfxsize\textwidth\epsfbox{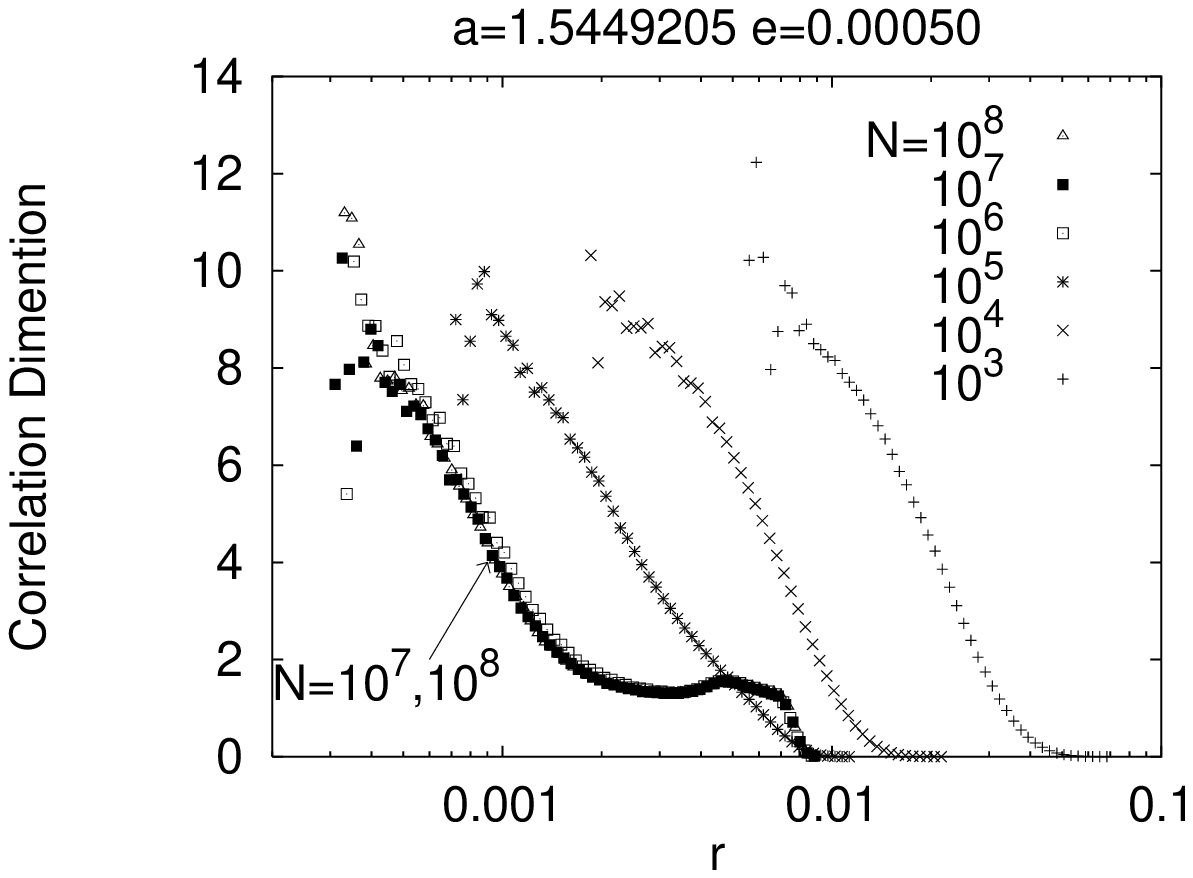}}
\caption{Correlation dimensions are plotted as a function of scale size
$r$ for different system size, which are indicated at the right of
each figure. The parameters are $a=1.699$, $\epsilon=0.008$(a), and
$a=1.5449205,\epsilon=0.00050$(b).}
\label{fig:CorDim}
\end{figure}

\begin{figure}
{\epsfxsize\textwidth\epsfbox{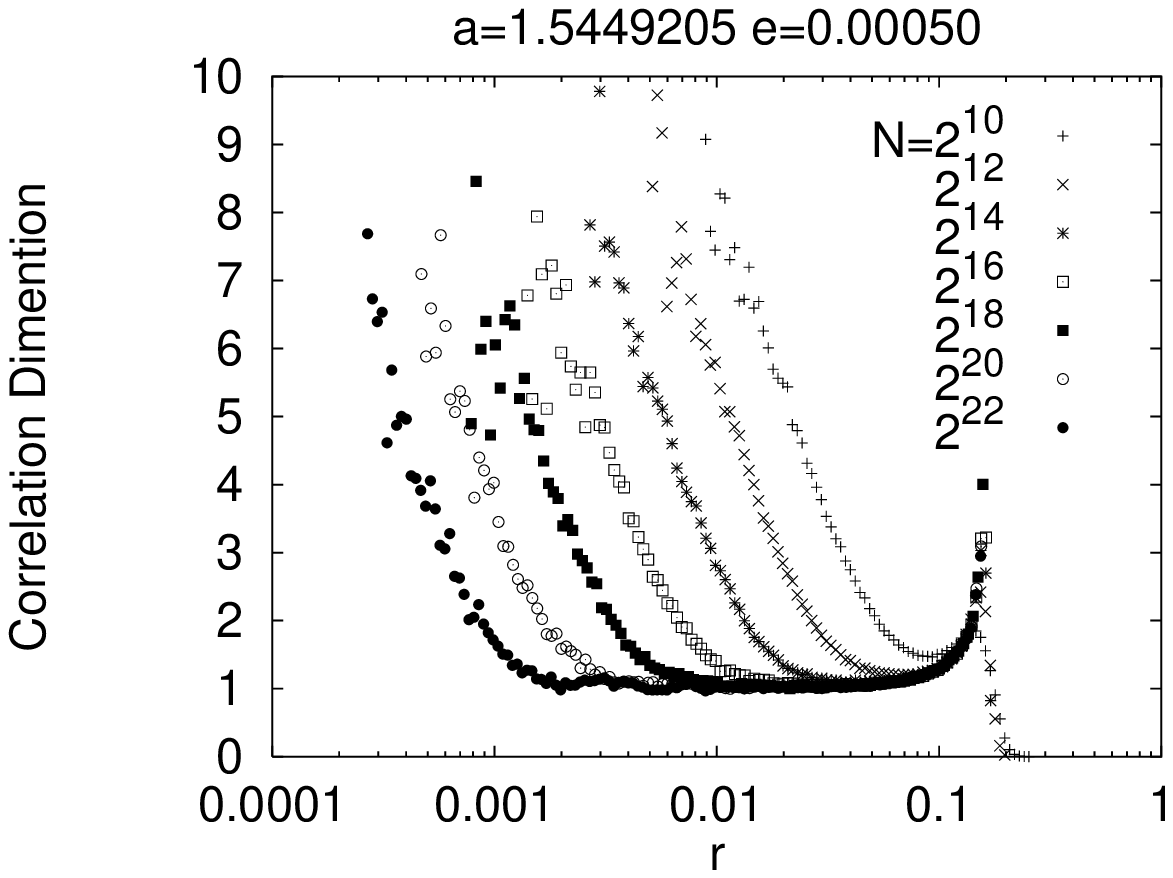}}
\caption{Correlation dimension for heterogeneous
GCM\protect\cite{Shibata1997}.  The parameters are $a=1.9,
\epsilon=0.11$. Nonlinearity parameter for each element is homogeneously
distributed over $a\in[1.875, 1.925]$.}
\label{fig:CorDimH}
\end{figure}

Next, we study the degrees of freedom in the collective dynamics in
the thermodynamic limit.  In the previous section, we have mentioned
that the collective motion, detected in the return map, has some
low-dimensional-like structure but the width of scattered points
around the `torus' remains finite in the thermodynamic limit.  Since
the possibility of higher dimensional torus is not excluded only by
the figure, we measure the correlation
dimension\cite{Grassberger-Procatia} of the mean-field time series.

In Fig.\ref{fig:CorDim}, the change of slope in the correlation
integral $\frac{d\log{C(r)}}{d\log{r}}$ is plotted as a function of
the scale size with increasing the system size $N$. For a smaller
system size, the correlation dimension is increased monotonically with
the decrease of the scale as in the case of random variable. For a
lager system size, curves have a plateau at a value less than the
correlation dimension two, which seems to correspond to the collective
motion. In a smaller scale, however, correlation dimension becomes
large.  At this smaller scale, the motion is hard to be
distinguishable from noise.  If the scale of this regime got smaller
with the size $N$, one could conclude that the collective dynamics is
low-dimensional in the thermodynamic limit.  As shown in
Fig.\ref{fig:CorDim}, this is not the case.  The slope function
converges to a certain curve with the increase of size $N$ where the
plateau region is no more expanded.  Thus, the mean field dynamics
does not converge to lower dimensional dynamics in the thermodynamic
limit.

To check the validity of this method, it will be relevant to mention
the case with a heterogeneous system\cite{Shibata1997}, e.g., a
globally coupled map with distributed nonlinearity parameter $a$ over
elements, where the mean-field dynamics shows a clear quasi-periodic
motion.  The width of scattered points around the tours converges to
$0$ in the thermodynamic limit. Corresponding plots of slopes are
given in Fig.\ref{fig:CorDimH}, where the plateau at the value 1 is
expanded with $N$, and the ``noise'' region is shrieked to the scale
$r\approx0$.  The difference from our uniform case is clearly
visible\footnote{In the heterogeneous case, the law of large numbers
might be considered to be recovered around the torus motion.}.

\subsection{Characteristic Time Scale of Collective Motion}

\begin{figure}
{\epsfxsize\textwidth\epsfbox{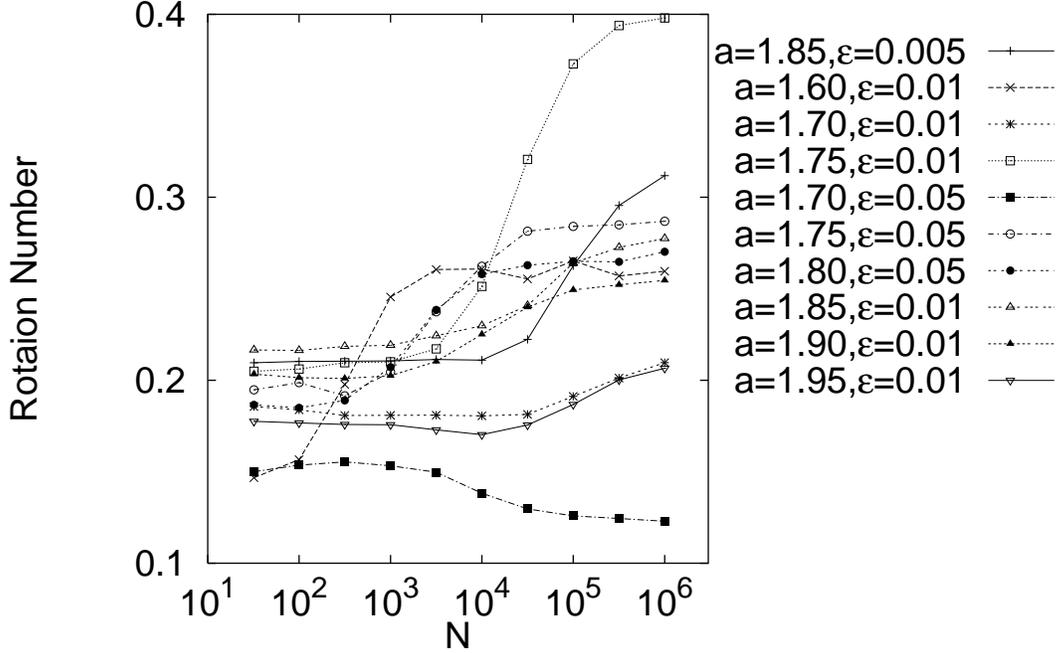}}
\caption{Rotation number of the mean field dynamics $R$, plotted as a
function of system size $N$.}
\label{fig:RotNum}
\end{figure}

\begin{figure}
{\epsfysize.3\textheight\epsfbox{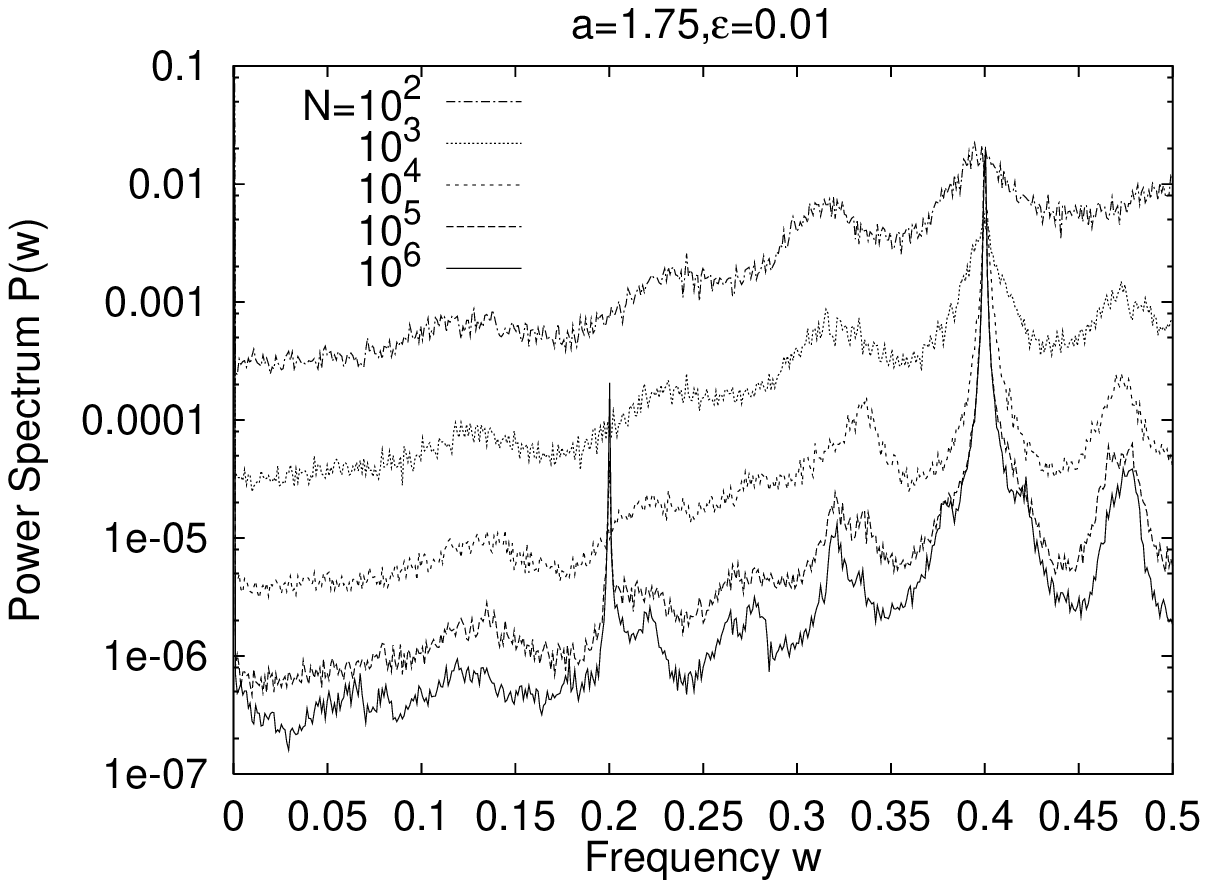}}
{\epsfysize.3\textheight\epsfbox{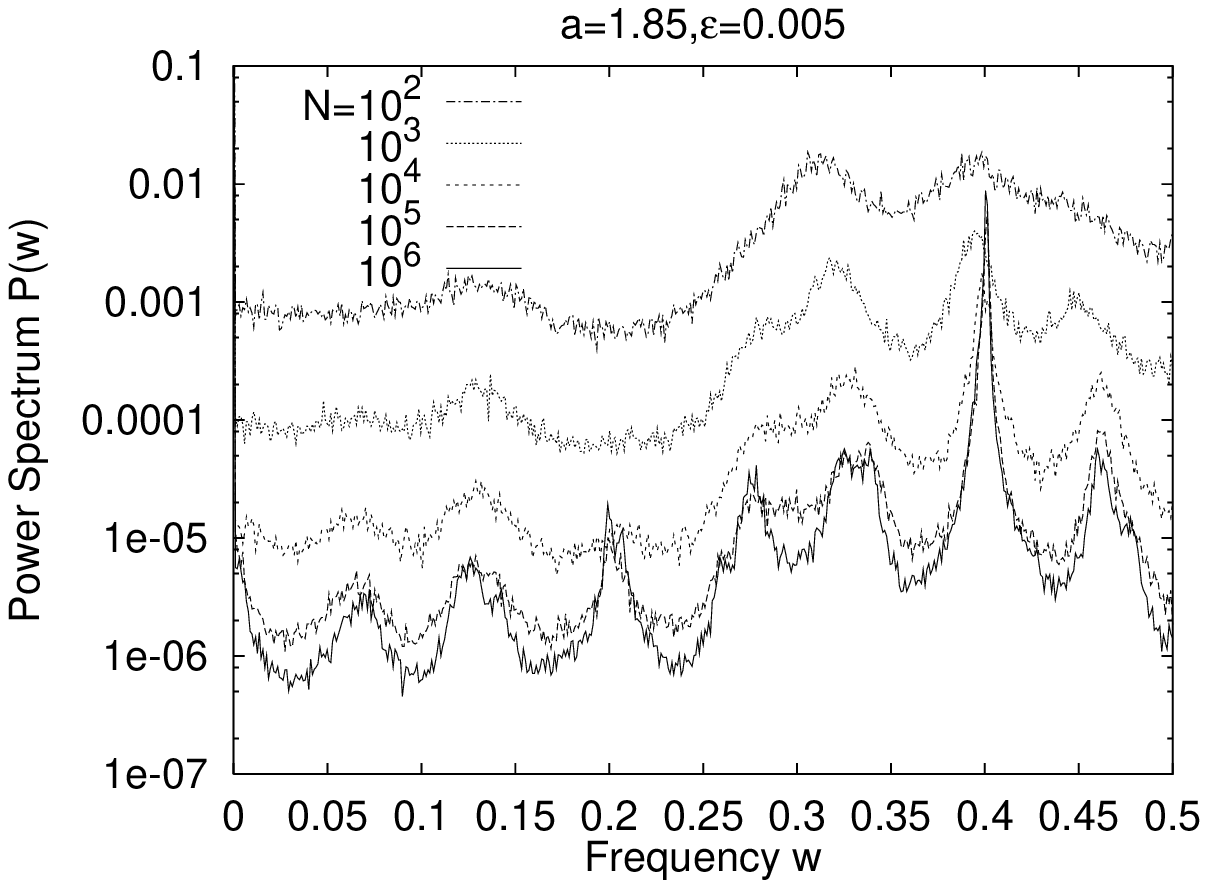}}
{\epsfysize.3\textheight\epsfbox{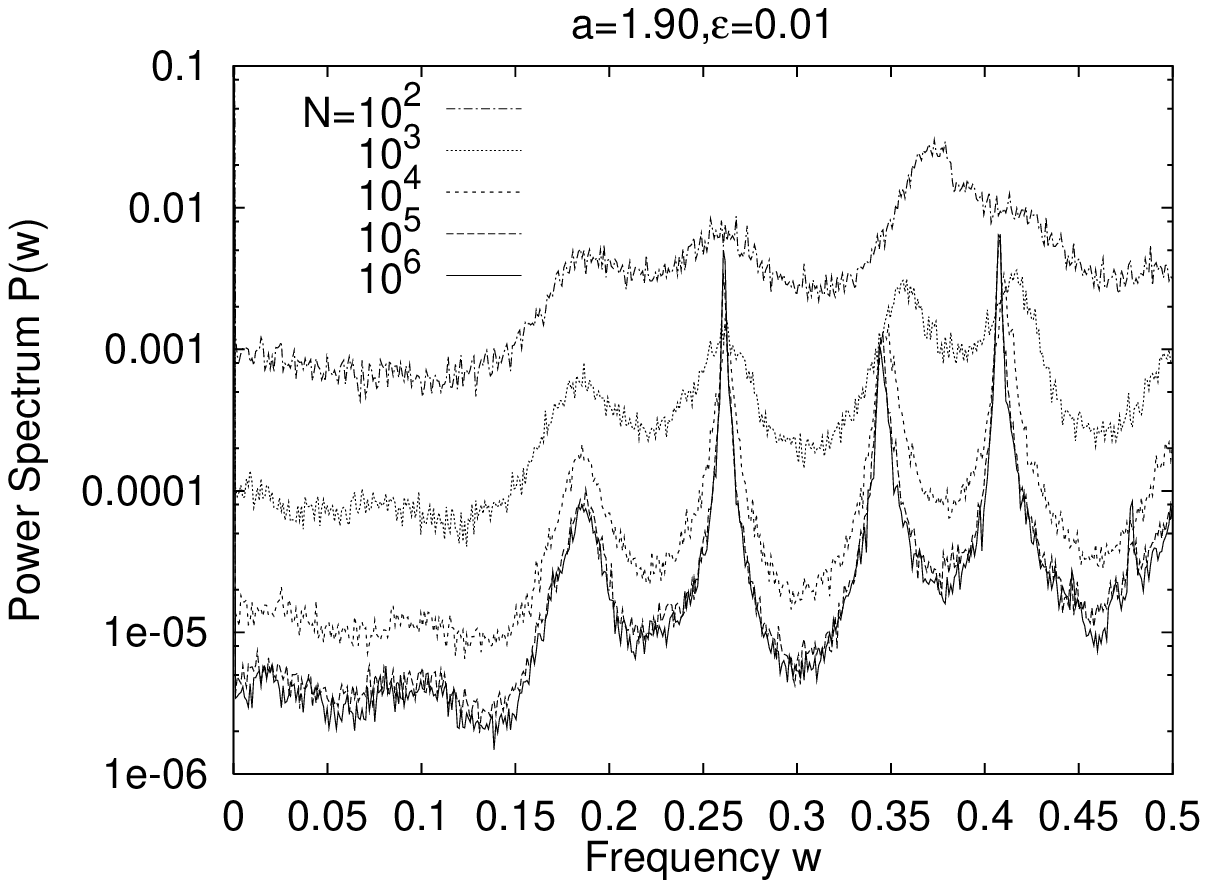}}
\caption{Power spectrum of the mean field dynamics with the increase
of system size $N$.}
\label{fig:PowervsN}
\end{figure}

To see how the time scale of the collective motion depends on the
system size $N$, we have measured the rotation number of the mean
field dynamics as a function of the system size $N$.  Here, the
rotation number $R$ is defined as
\begin{equation}
R=\lim_{t\rightarrow\infty}\frac{1}{t}%
\sum_{n=1}^{t}\frac{\Delta\theta_{n}}{2\pi} 
\end{equation}
where $\Delta\theta_{n}$ is angle variable formed by two vector
$(h_{n}-\langle h \rangle, h_{n+1}-\langle h \rangle)$,
$(h_{n+1}-\langle h \rangle, h_{n+2}-\langle h \rangle)$ defined
around the average mean filed $\langle h \rangle$ over time.

In Fig.\ref{fig:RotNum}, the rotation number converges to a certain
value.  This implies the appearance of characteristic time scale in
the mean field dynamics, independently of the system size for large
enough $N$.

Power spectrum of the mean field dynamics also supports the existence
of characteristic time scale of the mean field dynamics in the
thermodynamic limit as is shown in Fig.\ref{fig:PowervsN}. While the
spectra indicate that the mean field dynamics is non-periodic, there
are peaks, which get shaper with the increase of $N$ up to certain
size and converge to a certain curve.

\section{Global Phase Diagram of Collective Motion in Parameter Space;
Tongue-Like Bifurcation Structures}
\label{sec:PhaseDia}

As we have shown in the previous section(\S\ref{sec:phenomena}), the
mean-field dynamics depends on the parameter $a$, and $\epsilon$.  In
this section, we study how the collective motion depends on the
parameters.  To characterize this dependence on the parameters it is
often convenient to use the the mean square deviation(MSD) of the mean
field distribution as we have introduced in \S\ref{sec:thermo}.

\subsection{Phase Diagram in ($a, \epsilon$)-Plain}

\begin{figure}
{\epsfxsize\textwidth\epsfbox{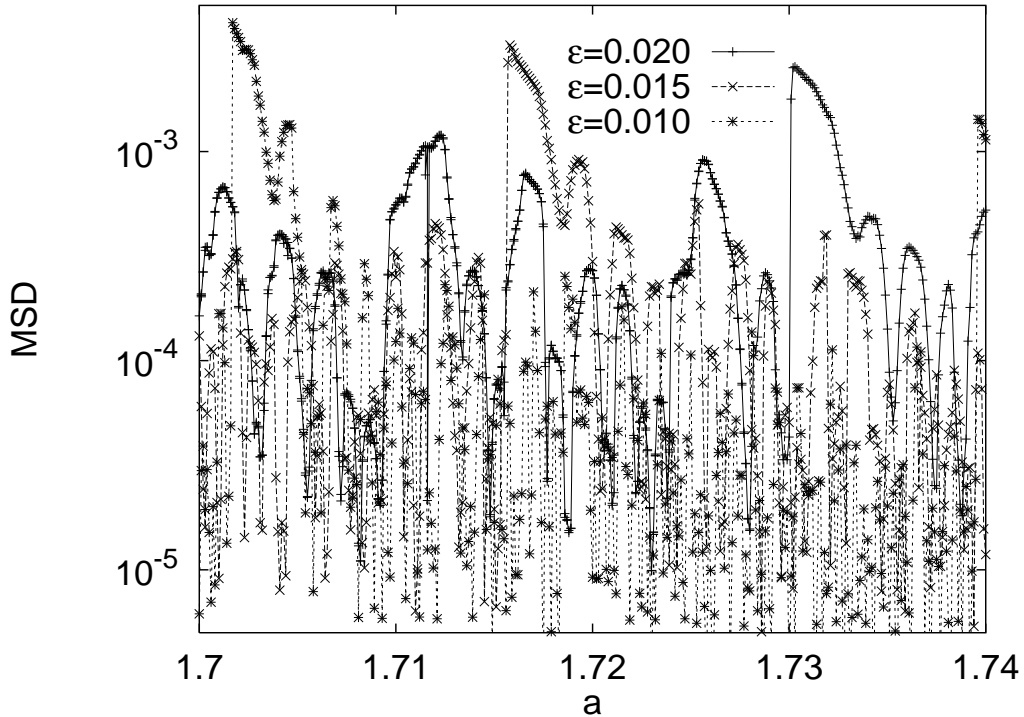}}
\caption{Mean square deviation (MSD) of the mean field dynamics is
plotted as a function of $a$. $\epsilon=0.02(\diamond), 0.15(+),
0.01(\Box)$.$N=2^{16}$.}
\label{fig:MSDvsa}
\end{figure}

\begin{figure}
{\epsfysize.9\textheight\epsfbox{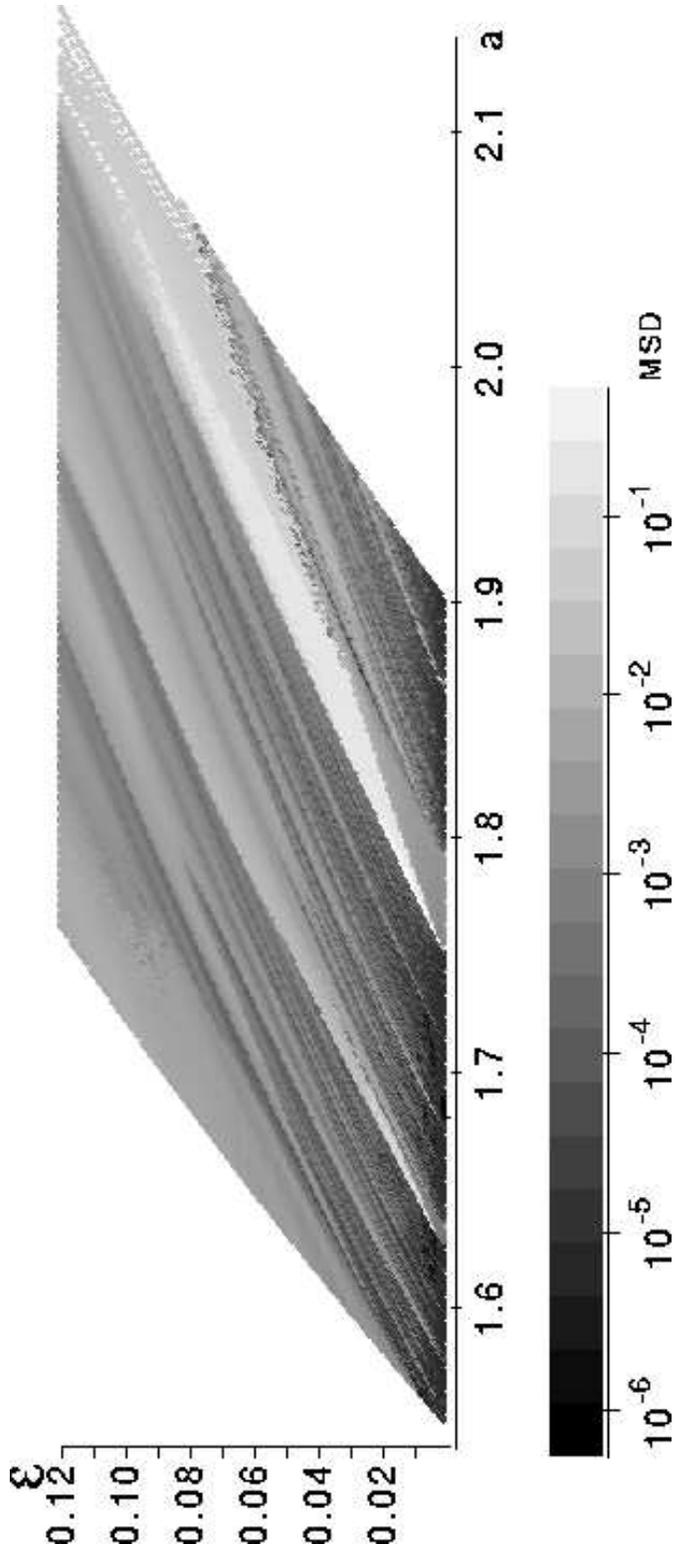}}
\caption{Mean square deviation (MSD) of the mean field dynamics is
plotted in $(a, \epsilon)$ plane with gray scale.  The scale shows the
value of MSD, where the darkest one corresponds to MSD$\approx
10^{-6}$, and the brightest one to MSD$\approx 10^{-1}$.}
\label{fig:MSD.noscale}
\end{figure}

In Fig.\ref{fig:MSDvsa}, the MSDs are plotted as functions of the
parameter $a$ for several coupling strengths $\epsilon$. Here the
system size is chosen to be large enough, to see the behavior of MSD
converged in the thermodynamic limit.  Two points should be noted
here.  First, the change of MSD is not monotonic with $a$, but is
rather complicated.  Second, the change of MSD is complicated with
fine structures, which still keep some similarity against the changes
of the coupling strength $\epsilon$.  For example, a similar but
slightly different structure is visible for $a\approx 1.7025$ for
$\epsilon=0.01$, $a\approx 1.725$ for $\epsilon=0.015$, and $a\approx
1.73$ for $\epsilon=0.02$.  In Fig.\ref{fig:MSD.noscale} the parameter
dependence of MSD is plotted on the 2-dimensional ($a,\epsilon$)
plane.  First, regimes with a larger collective motion form
tongue-like structures, each of which starts at some point or
intervals of parameter $a$ at $\epsilon=0$, and grows with $\epsilon$.
Second, the growth of the edge in a tongue-like structure has a
nonlinear dependence on the parameters $a$ and $\epsilon$.  Third, for
almost all parameter values, the MSD of the mean-field remains finite
in the thermodynamic limit.

\subsection{Effective Nonlinearity}
\label{ssec:Effective Nonlinearity}

\begin{figure}
{\epsfxsize\textwidth\epsfbox{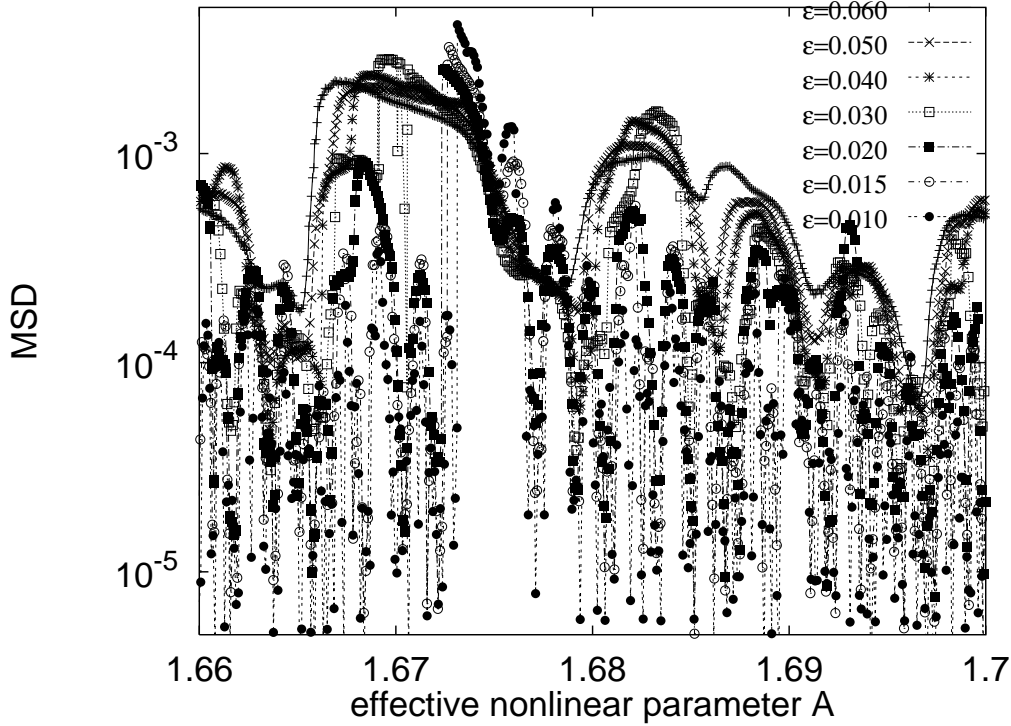}}
\caption{Mean square deviation (MSD) of the mean field dynamics $h_n$ are
plotted as functions of the effective nonlinearity parameter $A$.}
\label{fig:MSDvsA}
\end{figure}

\begin{figure}
{\epsfysize.85\textheight\epsfbox{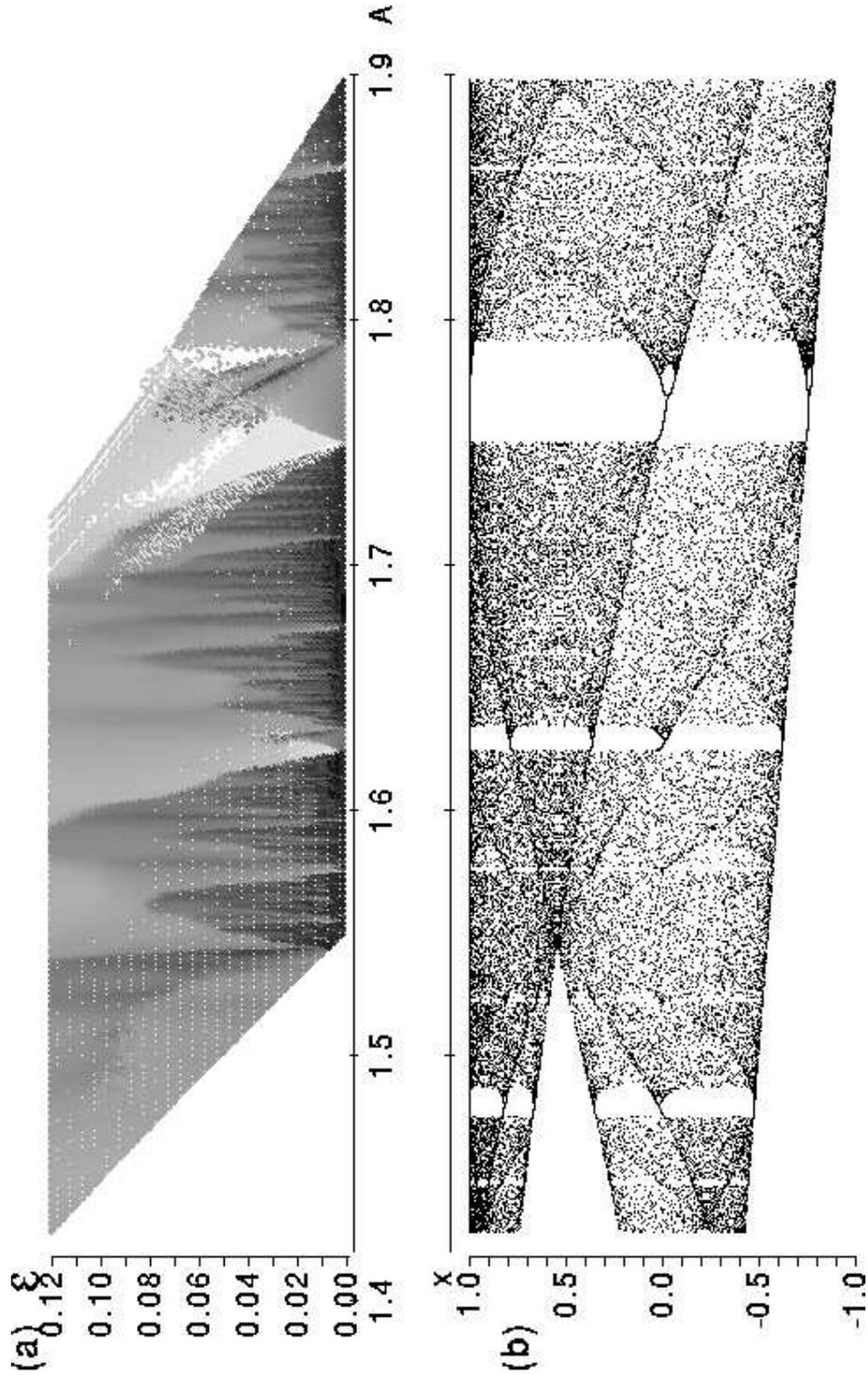}}
\caption{(a)Mean Square Deviation (MSD) of the mean field dynamics is
plotted as a function of the effective nonlinearity parameter $A$ and
$\epsilon$.  The scale shows the value of MSD, where the darkest one
corresponds to MSD$\approx 10^{-6}$, and the brightest one to
MSD$\approx 10^{-1}$ (see
Fig.\protect\ref{fig:MSD.noscale}).(b)Logistic map bifurcation diagram
with the increase of nonlinearity parameter. Horizontal axises are
common among two figures.}
\label{fig:MSD.scale}
\end{figure}

To see the structure in the parameter space closely, we introduce
rescaling of the parameters. For it, we note that each element obeys
the following dynamics,
\begin{equation}
x_{n+1}=(1-\epsilon)(1-a x_{n}^2)+\epsilon h_n
\end{equation}
where $h_n$ is the mean-filed value at time step $n$, which can be
considered as a time dependent input to each element.  In this map,
the nonlinearity is modified by the additional term $h_n$.  Taking
into account of this point, we normalize the variable $x_n$ as
follows:
\begin{equation}
x_n \rightarrow (1-\epsilon+\epsilon h_n)X_n.
\end{equation}
Then the logistic map of each element is written as
\begin{equation}
X_{n+1}=1-a(1-\epsilon) (1-\epsilon+\epsilon h_n){X_n}^2
\end{equation}
where $a(1-\epsilon) (1-\epsilon+\epsilon h_n)$ can be regarded as the
nonlinearity at time step $n$.

Since the above scaling is time-dependent due to $h_n$, we define the
effective nonlinearity parameter $A$ as
\begin{equation}
A=(1-\epsilon) (1-\epsilon+\epsilon \langle h \rangle)a.
\end{equation}
with the time independent rescaling of $x_n$,
\begin{equation}
x_n \rightarrow (1-\epsilon+\epsilon\langle h\rangle)X_n.
\end{equation}
where $\langle h \rangle$ is mean-field average in time,
\begin{equation}
\langle h \rangle=\lim_{t\rightarrow\infty}\frac{1}{t}\sum_{t=0}^{t}h_{n}.
\end{equation}

In Fig.\ref{fig:MSDvsA} we have plotted MSD by adopting the effective
nonlinearity parameter $A$ instead of $a$. In
Fig.\ref{fig:MSD.scale}(a) the parameter dependence of MSD is plotted
on the 2-dimensional ($A,\epsilon$)-plane. The scaling structure of
tongues seems to be much clearer. While the width of each tongue seems
to increase roughly linearly with $a$ and $\epsilon$, detailed
discussion will be appeared in \S\ref{sec:scaling}.

When the coupling strength $\epsilon$ approaches $0$, each tongue
structure corresponds to a window of the single logistic
map(Fig.\ref{fig:MSD.scale}(b)).  For instance, between $A\approx
1.75$ and $A\approx1.79$ a tongue structure is clearly shown in
Fig.\ref{fig:MSD.scale}(a), corresponding to the period-3 window of
the single logistic map.  Although there are countably infinite
windows in the parameter space in the logistic map, it is difficult to
detect the windows for a longer period numerically.  However, it is
remarkable that a lot of tongue structures are visible in our model,
corresponding to the windows with a longer period.

In each tongue structure, further internal structures exist.  For
instance, the tongue corresponding to period-3 window of the logistic
map between $A=1.75$ and $A\approx1.79$, has three internal
structures, roughly speaking. To understand each inner structure in
the tongue, we will study the dynamics of each element and the
distribution in the following sections(\S\ref{sec:selfconsistent}, and 
\S\ref{sec:bif}).

\section{Collective Behavior Through Self-Consistent Dynamics}
\label{sec:selfconsistent}

In this section, we briefly describe how the collective motion is
formed, especially focusing on the tongue structure.

\subsection{Distribution Dynamics}

\begin{figure}
{\epsfxsize.8\textwidth\epsfbox{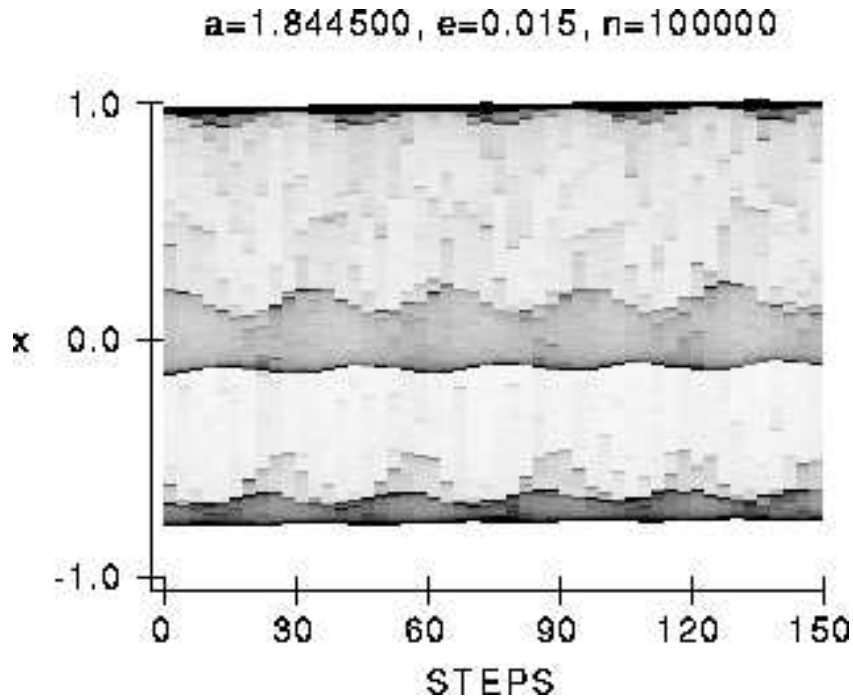}}
\caption{The distribution dynamics is plotted as a function of
time. The density of the population is shown with the use of a gray
scale. The darker region indicates the high density. The parameters
are $a=1.8445$,$\epsilon=0.015$, $N=10^5$. }
\label{fig:DistDyn}
\end{figure}

In the limit of $N\rightarrow\infty$, the probability distribution
function is defined as follows,
\begin{equation}
\rho_{n}(x) = 
\lim_{N\rightarrow\infty}\frac{1}{N}\sum_{i}\delta(x-x_{n}(i)).
\end{equation}
Oscillation, rather than the fixed point, of the mean field dynamics
implies that the probability distribution function does not also
remain stationary but depends on time.

Time series of the probability distribution function by numerical
calculation is given in Fig.\ref{fig:DistDyn}.  The parameters for the
figure $(a=1.8445, \epsilon=0.015)$ belong to the tongue structure in
the period 3 window. In this case, since the mean field dynamics has
the component of period 3, we plot the figure by every 3 steps to see
the slow modulation of $\rho_{n}(x)$.  Due to the chaotic oscillation
of each element, the distribution function spreads over $x\in[-0.8,
1.0]$, but the distribution is not monotonous, and has some structure.
In the three regions around $x\approx 1.0, 0.0, -0.8$, the population
are relatively large.  This number ``three'' corresponds to the period
of window which the tongue structure of this motion corresponds to.
The number of elements in each three region oscillates in time, and
the phase of each oscillation is different from that of the others.

\subsection{Formation of Self-Consistent Dynamics}
\label{ssec:selfconsis}

\begin{figure}
{\epsfxsize\textwidth\epsfbox{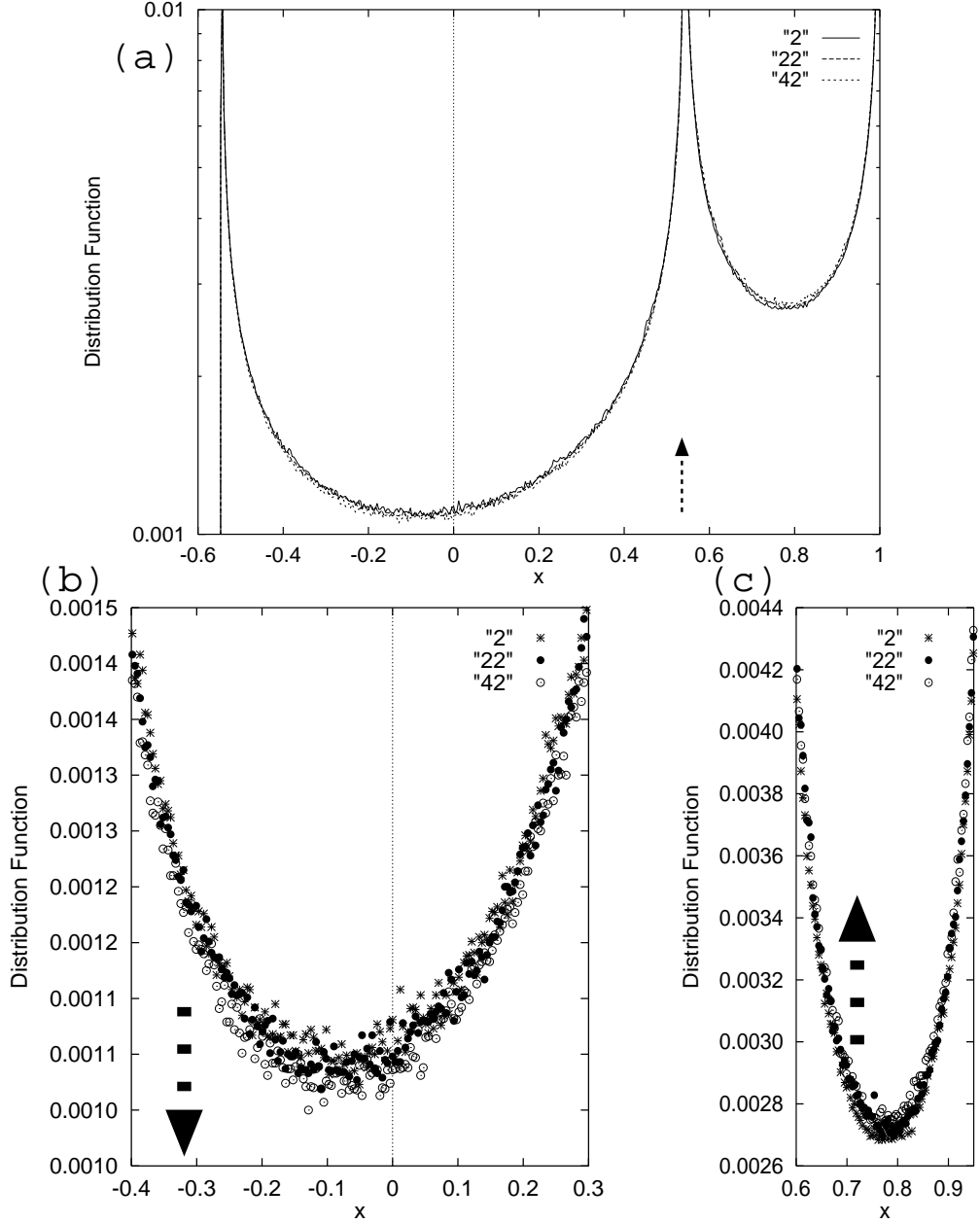}}
\caption{The distribution functions at time $n=0(\ast)$, $20(\bullet)$,
$40(\circ)$ are shown.}
\label{fig:DisFunc}
\end{figure}

\begin{figure}
{\epsfxsize\textwidth\epsfbox{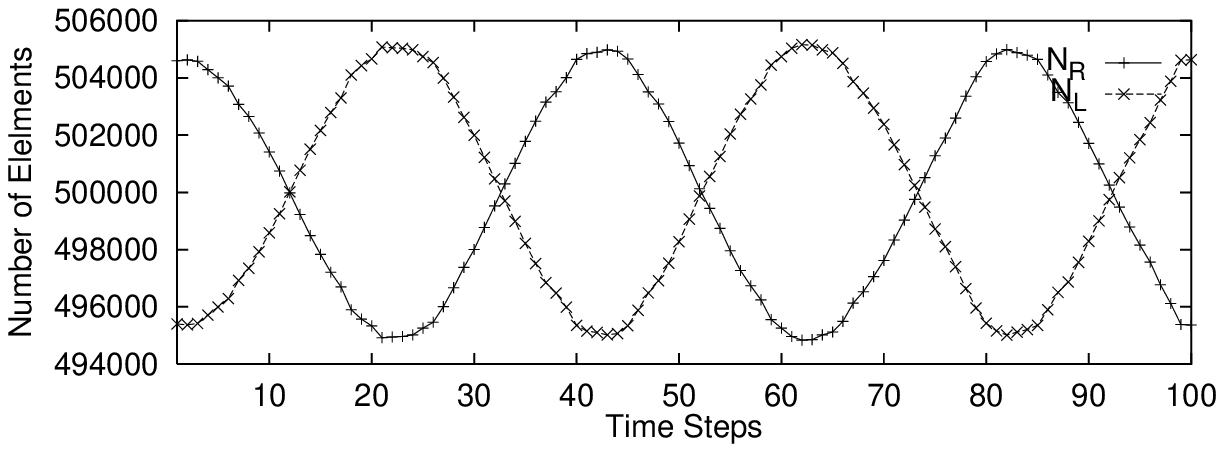}}
{\epsfxsize\textwidth\epsfbox{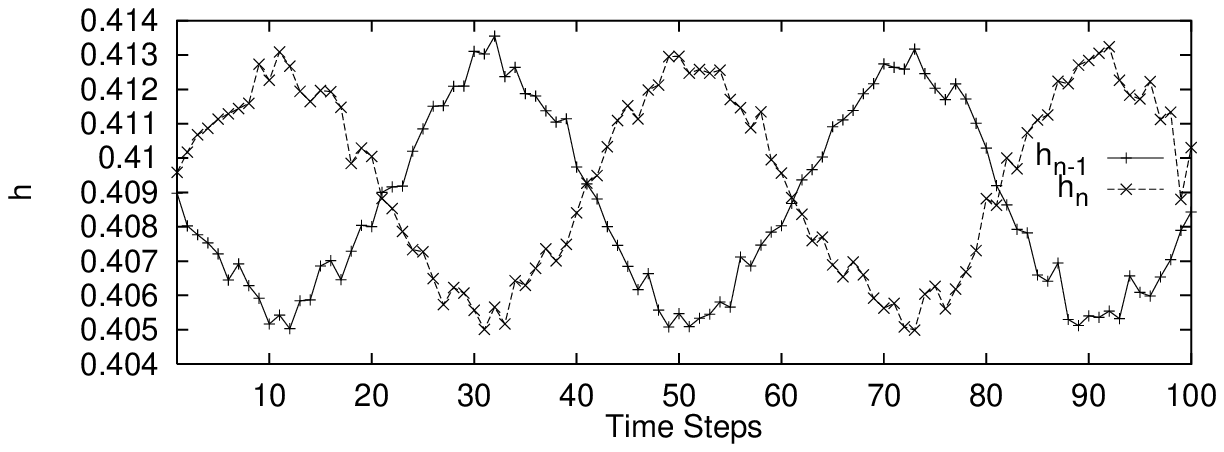}}
\caption{(a)Time series of the number of elements in the 2 regions,
(b)time series of the mean field.$a=1.5449205,
\epsilon=0.0005, N=10^6$. }
\label{fig:SelfCon}
\end{figure}

\begin{figure}
{\epsfxsize.7\textwidth\epsfbox{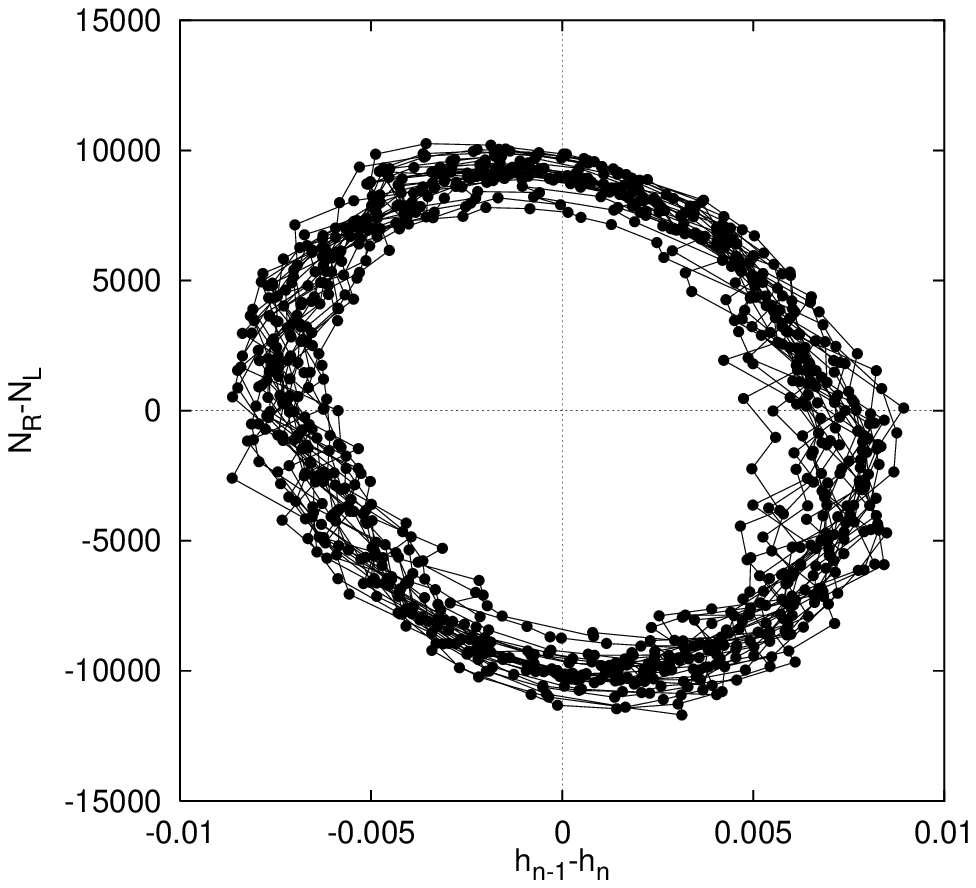}}
\caption{Return map of the time series of number of elements in the 2
regions and time series of the mean field, plotted per two steps for
even $n$ and odd $n$.  $a=1.5449205,
\epsilon=0.0005, N=10^6$.}
\label{fig:RMSelfCon}
\end{figure}

It is interesting to view our collective dynamics as an interference
between mean field dynamics and individual elements. Before we present
a scenario for collective motion, we demonstrate the formation of
self-consistent dynamics between the mean field dynamics and
individual elements as follows.

For simplicity, we adopt the case, in which the effective nonlinearity
parameter $A$ is near the period-2 band merging point(the parameter
are $a=1.5449205$, $\epsilon=0.0005$, and the time series and the
return map for the parameter are given in Fig.\ref{fig:rm.p2}). The
distribution function is given in Fig.\ref{fig:DisFunc} at every 20
steps. In this case, distribution of elements can be divided into two
regions around $x\approx 0.54$. During these 40 time steps the value
of distribution function at the left region ($x<0.54$), given in
Fig.\ref{fig:DisFunc}(b), decreases with time, while the other region
plotted in Fig.\ref{fig:DisFunc}(c) increases with time. Although the
change of distribution is quite small, there is a systematic
oscillation in the distribution(cf.Fig.\ref{fig:hist}).

Consider a population dynamics of each of the two regions. In
Fig.\ref{fig:SelfCon}(a), the population in each region, $N_{L}$, and
$N_{R}$ are plotted as a function of time.  $N_{L}$ denotes the number
of elements in the region smaller than $x\cong0.54$ in
Fig.\ref{fig:DisFunc}, while $N_{R}(=N-N_{L})$ denotes that for larger
than $x\cong0.54$. (The definition for each region is described below
in detail.)  The population in each region oscillates in time.  In
Fig.\ref{fig:SelfCon}(b), on the other hand, since the mean field has
a component of period two, the time series $h_{n-1}$ and $h_{n}$ are
plotted at every two steps. Note that the mean field also oscillates in
time with the same period as $N_R$, and $N_L$, but the phase of the
mean field oscillation is different from that of the population
dynamics in Fig.\ref{fig:SelfCon}(a).

To see how the mean filed dynamics and the distribution dynamics
interfere each other, we construct a return map of above two
quantities.  Fig.\ref{fig:RMSelfCon} is a return map of the
distribution dynamics and the mean field dynamics.  This figure
implies that a self-consistent dynamics is formed as follows,
\begin{equation}
\left\{
\begin{array}{lll}
\tilde{h}_{n}&=&\tilde{h}(\tilde{h}_{n-1},\tilde{N}_{n-1}),\\
\tilde{N}_{n}&=&\tilde{N}(\tilde{h}_{n-1},\tilde{N}_{n-1}),
\end{array}
\right.
\end{equation}
where each $\tilde{h}$ and $\tilde{N}$ is a function of
$\tilde{h}_{n}=h_{n-1}-h_{n}$, and $\tilde{N}_{n}=N_{L}-N_{R}$.  If
the mean field were an external force for each element, the population
responds to the mean field value.  On the other hand, since population
organizes the mean filed dynamics, the collective motion can be
described as a self-consistent relation between the population
dynamics and the mean field dynamics.

From the above viewpoint, we now demonstrate how the population
distribution is modified as the mean field varies slowly.

\subsection{Internal Bifurcation in Temporal Domain}
\label{ssec:int}

\begin{figure}
{\epsfxsize.45\textwidth\epsfbox{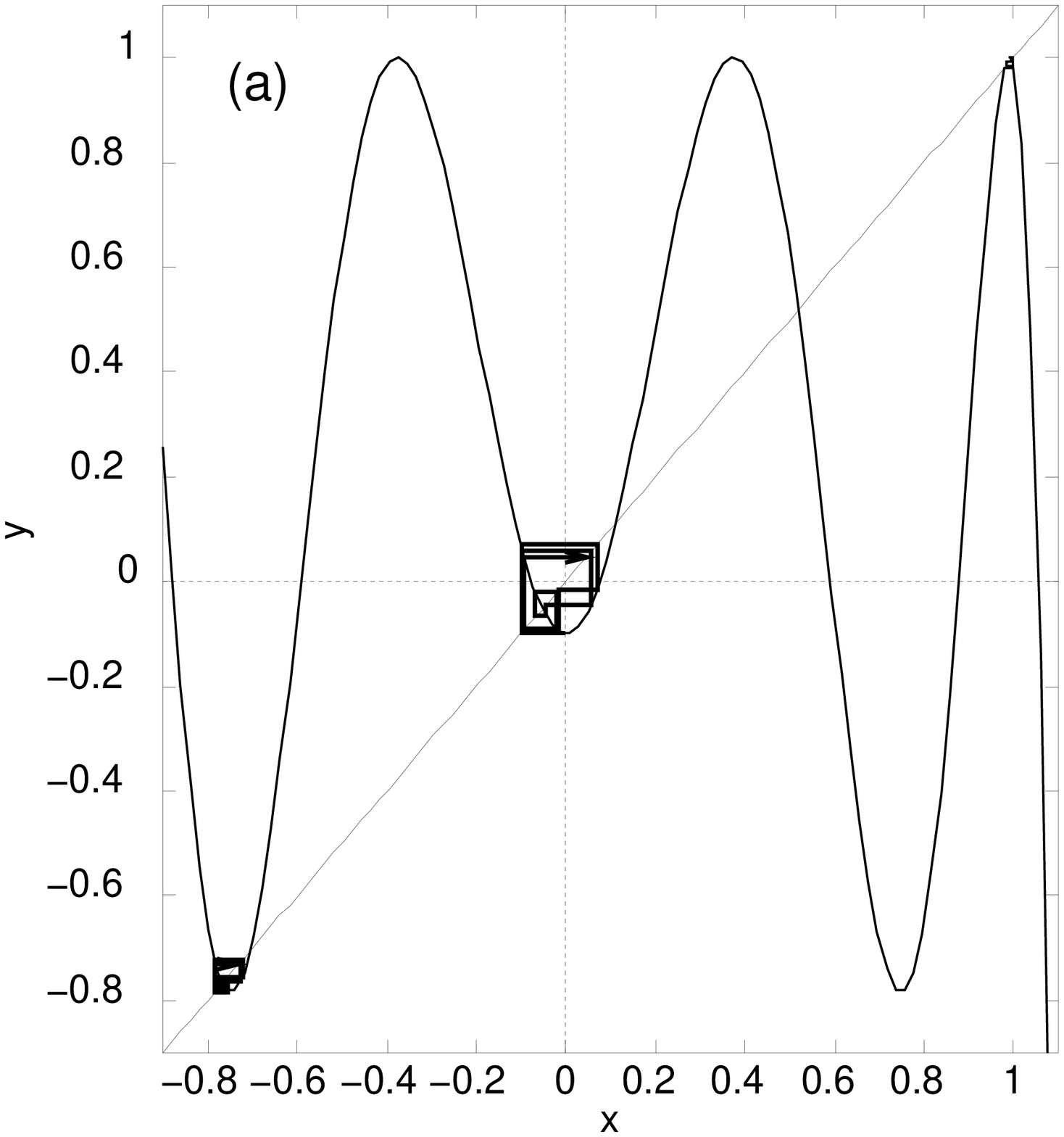}}
{\epsfxsize.45\textwidth\epsfbox{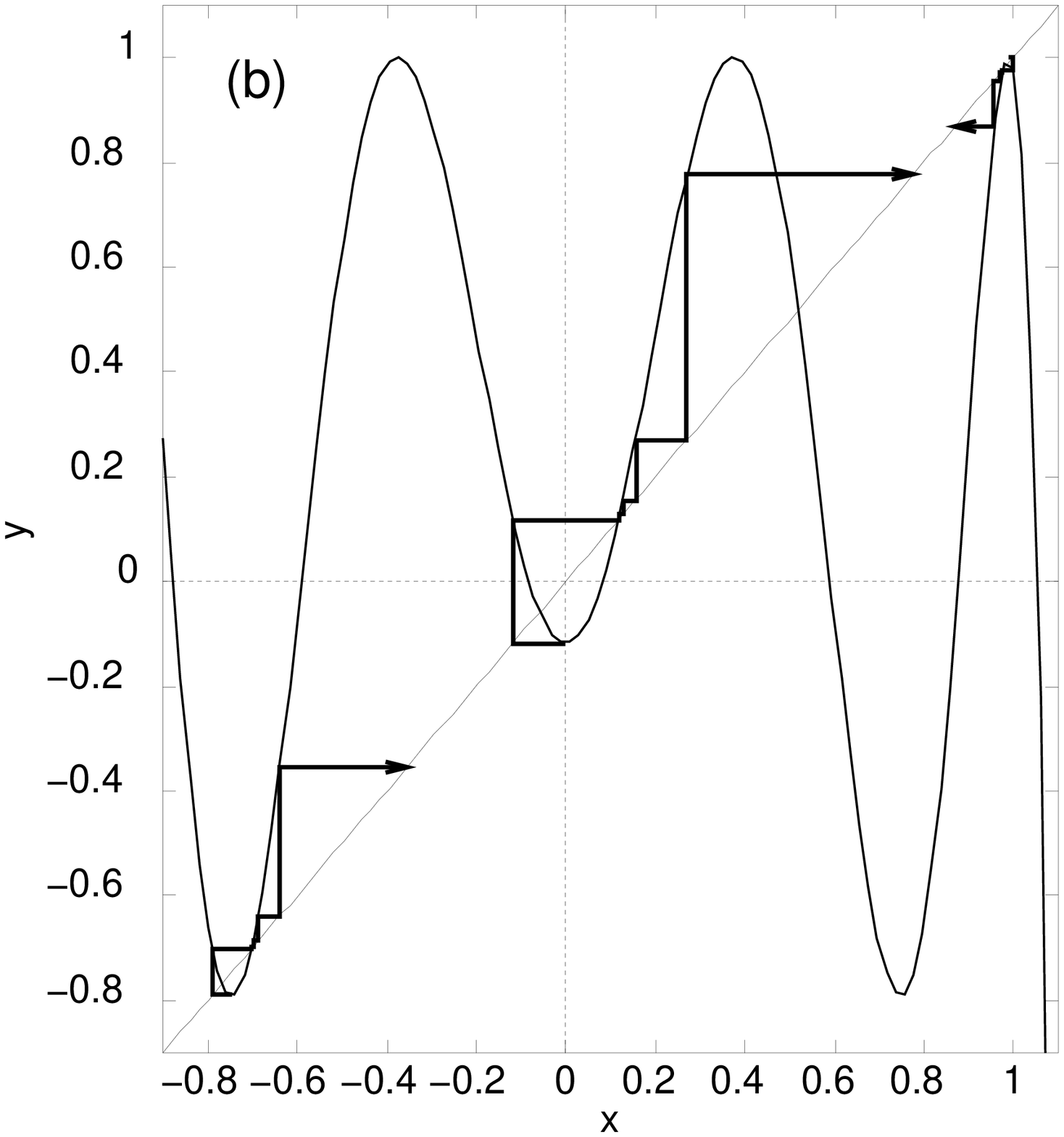}}
\caption{Behavior of the third iterate of a logistic map. (a) After tangent
bifurcation at three points, and (b)after crisis. Solid lines with
arrow indicate iterated trajectories starting from three points}
\label{fig:logistic2}
\end{figure}

\begin{figure}
{\epsfxsize.45\textwidth\epsfbox{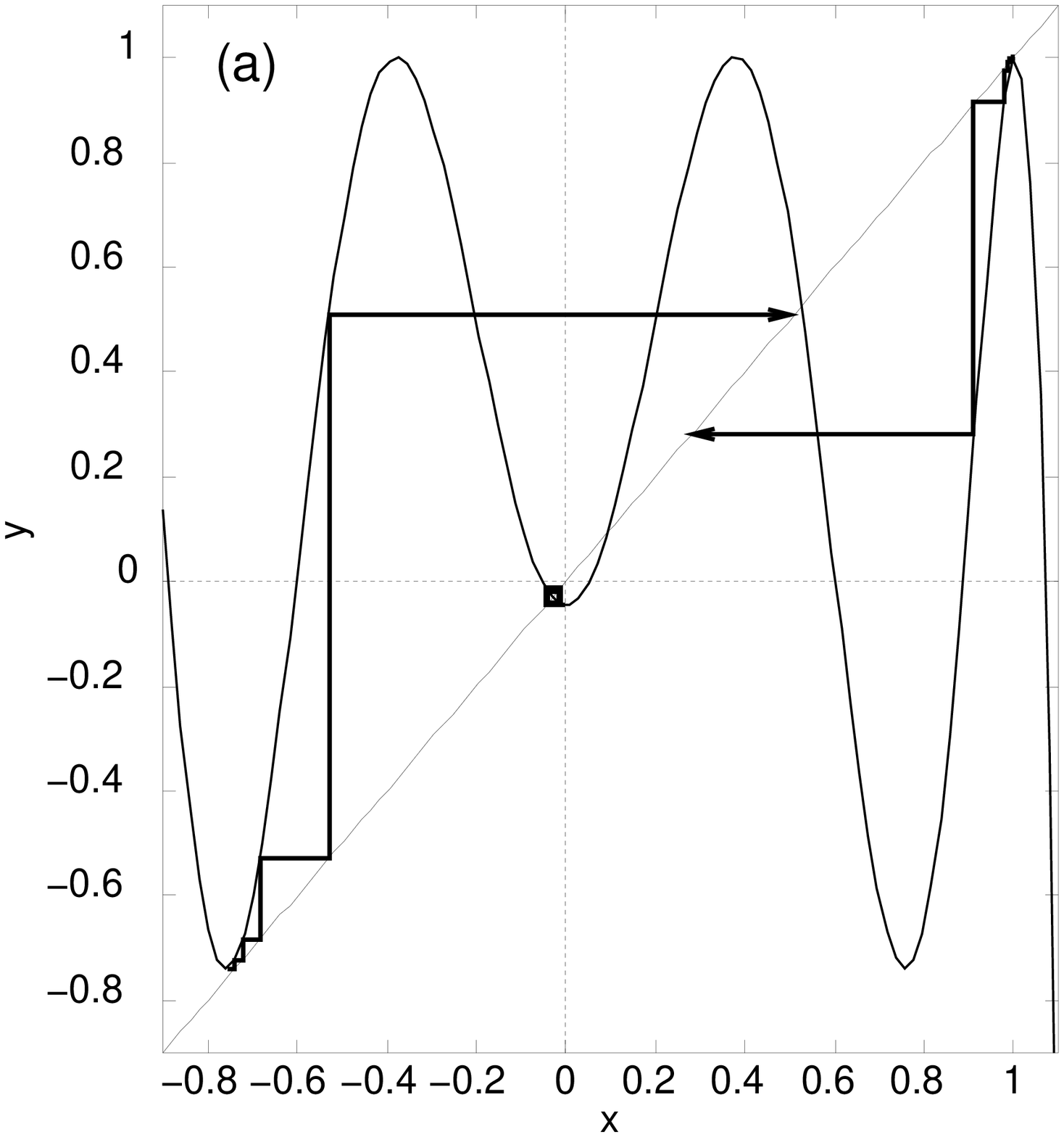}}
{\epsfxsize.45\textwidth\epsfbox{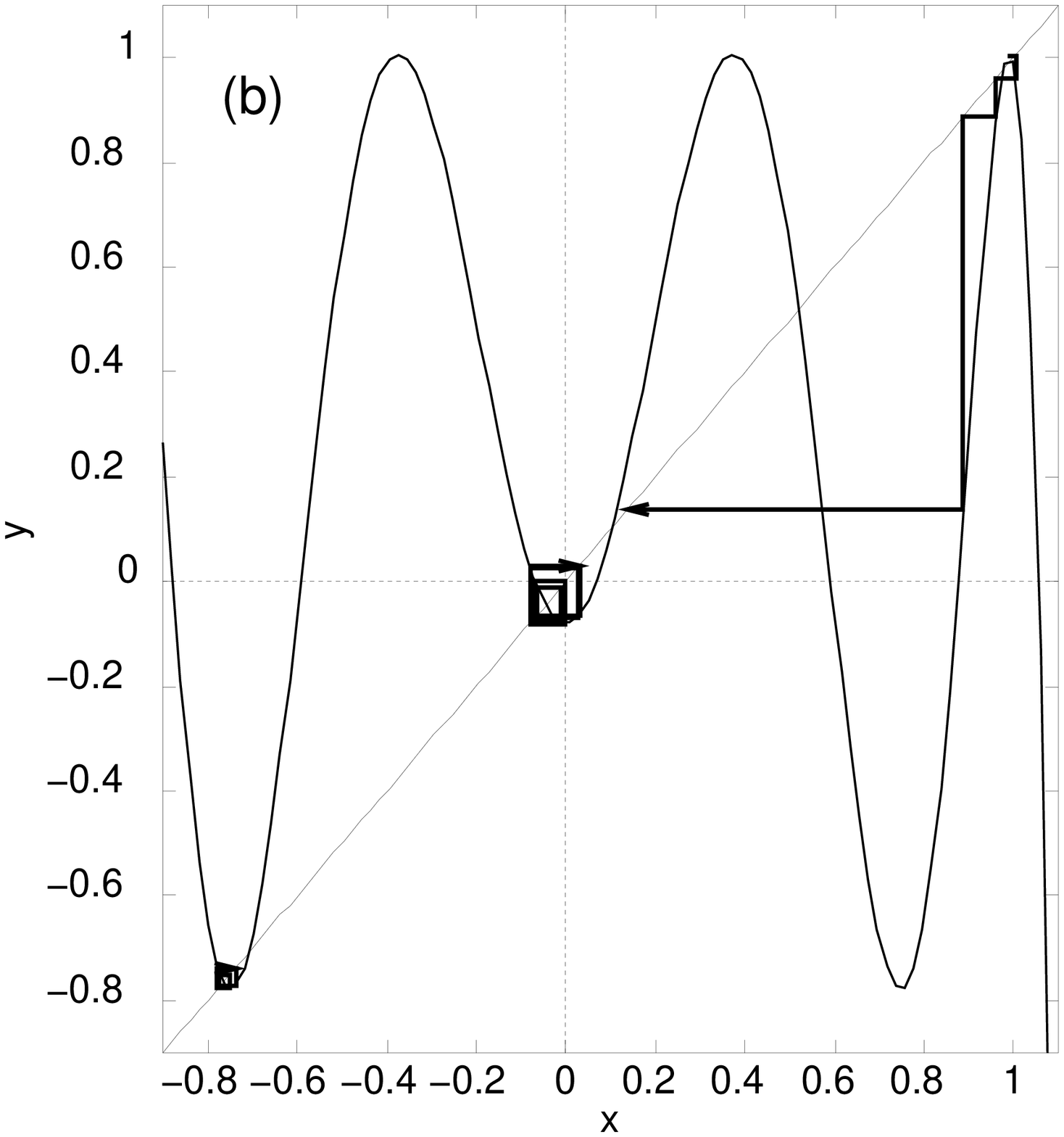}}
{\epsfxsize.45\textwidth\epsfbox{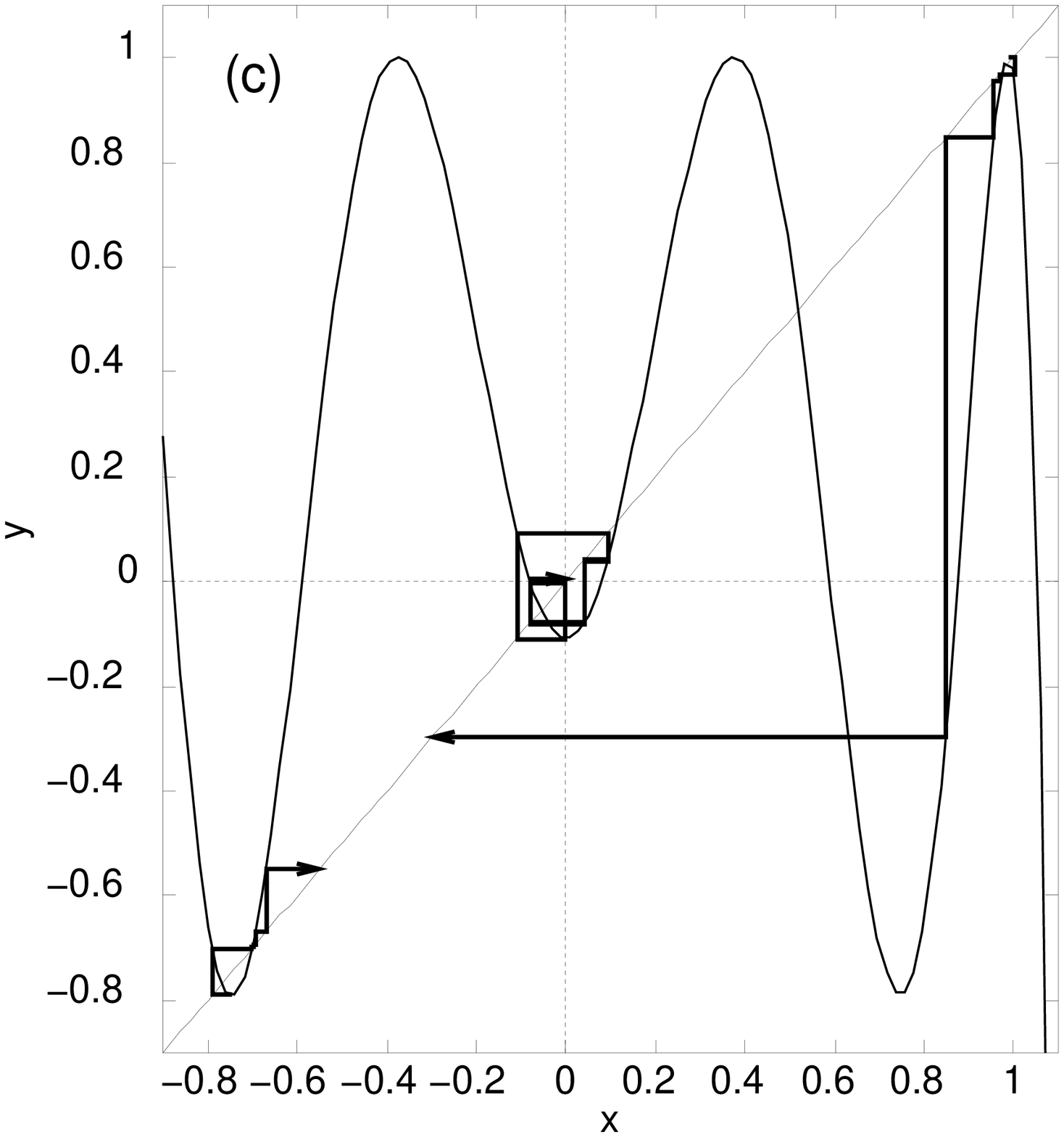}}
\caption{Behavior of the third iterate of the map 
Eq.(\protect\ref{eq:inteq}). In contrast with the case of
Fig.\protect\ref{fig:logistic2}, only one or two regions can attract
trajectories.  (a)The region around $x=0$ attracts orbits as a region
after tangent bifurcation, while the other regions, which are before
the tangent bifurcation, cannot attract orbits.  (b)Two regions around
$x=0$, and $x=-0.8$ attract orbits, while the region around $x=1.0$
can not attract orbits due to the crisis.  (c)The region around $x=0$
attracts orbits, while the other regions can not attract orbits due to
the the crisis.}
\label{fig:logistic3}
\end{figure}

\begin{figure}
{\epsfxsize\textwidth\epsfbox{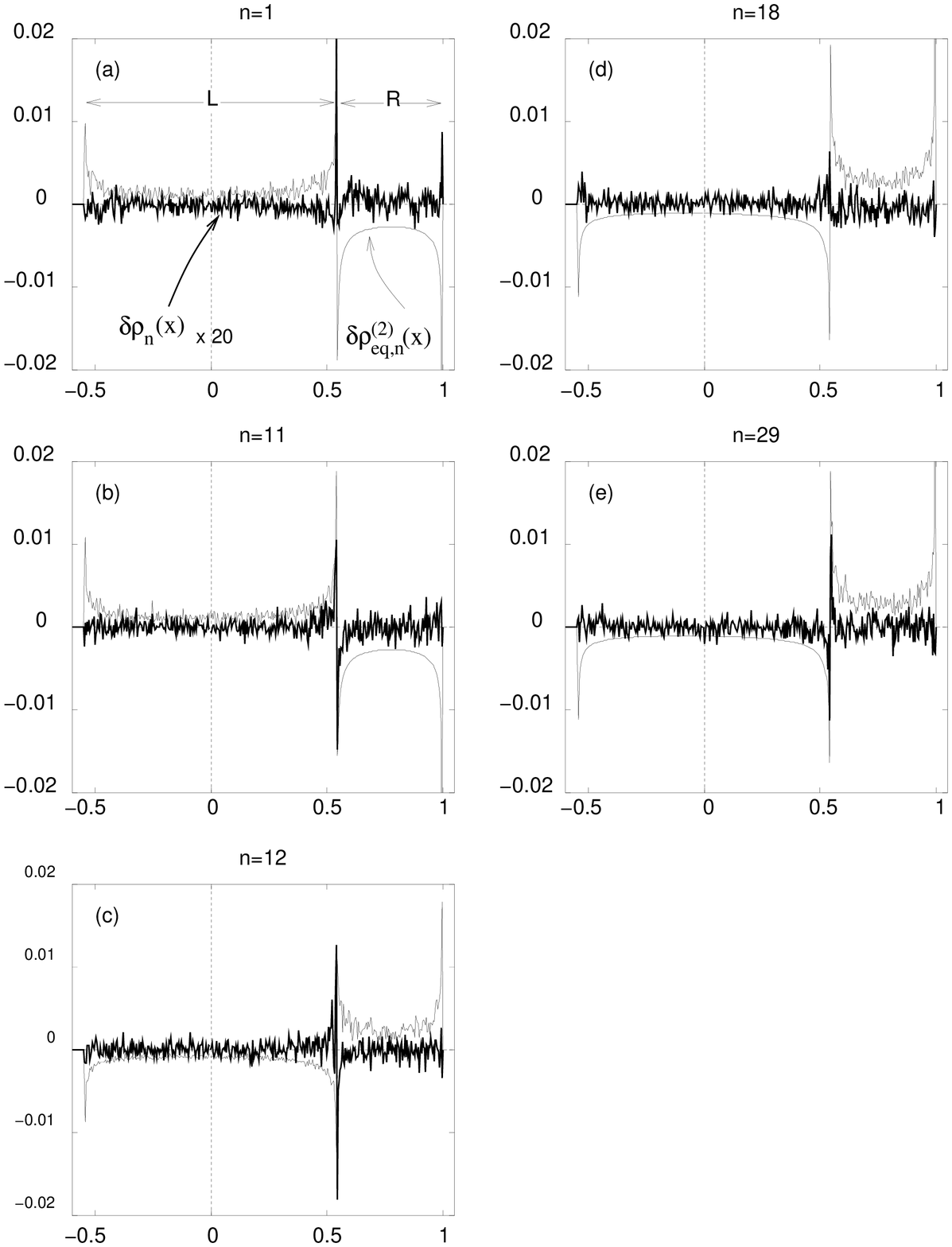}}
\caption{Temporal change of 
$\delta\rho_n(x)=\rho_{n}(x)-\langle\rho(x)\rangle$ is plotted by the
thick line by magnifying the scale 20 times, while
$\delta\rho_{eq,n}^{(2)}(x)=\rho_{eq,n}^{(2)}(x)-\langle\rho(x)\rangle$
is plotted by the thin line, where $\langle\rho(x)\rangle=
\lim_{t\rightarrow\infty}\frac{1}{t}\sum_{n=1}^{t}\rho_{n}(x)$.
We have plotted them instead of $\rho_{n}(x)$ and
$\rho_{eq,n}^{(2)}(x)$ whose change is too small to be visible.
Roughly speaking, the population in the region where
$\delta\rho_{eq,n}^{(2)}(x)$ is negative is going to the region where
$\delta\rho_{eq,n}^{(2)}(x)$ is positive.
$\delta\rho_{eq,n}^{(2)}(x)$ changes qualitatively at $n=12$ (c).
$a=1.5449176,\epsilon=0.0005,N=10^6$.}
\label{fig:hist}
\end{figure}

If the mean field were an external force for each element, it could be
regarded that each element follows the logistic map with external
force. This is valid if the mean field varies slowly.  In this case,
the motion of equation for the $i$'th element is given by
\begin{equation}
x_{n+1}(i)=F_{n}(x_n(i)),
\end{equation}
with
\begin{equation}
F_{n}(x)=(1-\epsilon)(1-ax^2)
+\epsilon\langle h \rangle +\epsilon\cdot\delta h_n,
\label{eq:inteq}
\end{equation}
where $\delta h_n$ is difference from $\langle h \rangle$, i.e.,
$h_n=\langle h\rangle + \delta h_n$.

If we took $\delta h_n$ out of consideration, the dynamics of each
element would be same as the logistic map with the nonlinearity
parameter $A=a(1-\epsilon)(1-\epsilon+\epsilon \langle h \rangle)$.
Since in the previous section (\S\ref{sec:PhaseDia}) each tongue
structure has good correspondence with a window of the logistic map,
here we especially focus on the window structure of the logistic map.
In the case of the logistic map, the period $p$ window starts at the
tangent bifurcation point of the $p$'th iterate of the map, and then
the period doubling bifurcation proceeds with the increase of $a$,
until the window ends up by crisis(see Fig.\ref{fig:logistic2}).  Note
that for a period $p$ window tangent bifurcation or crisis occurs at
$p$ points of $x$ at the same value of $A$. In this case, since
Eq.(\ref{eq:inteq}) has an invariant measure, the probability
distribution in each of $p$ pieces of regions is equivalent.

Take $\delta h_n$ in Eq.(\ref{eq:inteq}) into account as an external
force.  The bifurcation of a logistic map with time dependent external
force has a crucial difference from usual bifurcation of the logistic
map. In Fig.\ref{fig:logistic3}, examples of the third iterates of the
map with external force are shown. In Fig.\ref{fig:logistic3}(a) a
region around $x\approx 0$ crosses $y=x$, while two regions around
$x\approx 0.95$ and $x\approx-0.75$ do not cross $y=x$.  In
Fig.\ref{fig:logistic3}(b)(c), while three regions cross $y=x$, one or
two of the regions are collapsed by crisis. 

In general, consider the case of period $p$ window with external
force.  Since the tangent bifurcation or crisis of $p$ points occurs
at a different value of $A$, the number of divided attractors can be
changed with $A$.  Even if trajectories are attracted into $p$
distinct regions, the probability distribution in each region is not
equivalent.

As we have seen in the previous subsection(\S\ref{ssec:selfconsis}),
slow modulation of the mean field leads to the dynamics of the
distribution of population.  With the slow modulation of $\delta h_n$
in time, behavior of each element also changes.  In other words, with
the change of $\delta h_n$, bifurcation can occur in the effective map
for each element,
\begin{equation}
F^{(p)}_{n}=F_{n+p-1}\circ F_{n+p-2}\circ\cdots\circ F_{n+1}\circ
 F_{n},\label{eq:EffMap}
\end{equation}
which is the $p$'th iterate of the map Eq.(\ref{eq:inteq}).  Since
$\delta h_n$ changes temporally, such bifurcation occurs
temporally. To distinguish from the notion of bifurcation in parameter
space, such bifurcation is called as ``internal
bifurcation''\footnote{In our previous work
\cite{Shibata1997}, the nonlinearity parameter $a$ was distributed over
elements. In that case, some sort of differentiation of dynamics over
elements enabled the collective motion possible.  To characterize the
differentiation, the notion of ``internal bifurcation'' was introduced
as a snapshot representations of one system. As we will show below,
since the temporal bifurcation in an element leads to the collective
motion, we extend the notion of ``internal bifurcation'' to identical
case.}. While the notion of ``internal bifurcation'' indicates the
slow modulation of the effective map Eq.(\ref{eq:EffMap}), since time
dependence of $\delta h$ induces asymmetry in this effective map, as
we have shown in the previous paragraph, the dynamics of the mean
field has component of period $p$ motion. This is why we have plotted
the time series of the mean field and the probability distribution
function at every $p$ steps.

To characterize the effective map at every $p$ time steps, we
introduce the invariant measure $\rho_{eq,n}^{(p)}(x)$ determined from 
the Perron-Frobenius equation,
\begin{equation}
\rho_{eq,n}^{(p)}(x)
=\int_{-1}^{1}\rho_{eq,n}^{(p)}(y)\delta(x-F_{n}^{(p)}(y)))dy.
\label{eq:PF}
\end{equation}
If the mean field changes slowly, we can approximately regard our GCM
dynamics as relaxation process of $\rho_{n}(x)$ to
$\rho_{eq,n}^{(p)}(x)$.  In each time step, elements follow the
effective map Eq.(\ref{eq:EffMap}) so that the distribution function
$\rho_{n}(x)$ is going to be relaxed.  The elements in the region
where $\rho_{eq,n}^{(p)}(x)-\rho_{n}(x)<0$ are going to the region
where $\rho_{eq,n}^{(p)}(x)-\rho_{n}(x)>0$.  As a result $\rho_{n}(x)$
is going to relax into $\rho_{eq,n}^{(p)}(x)$.

On the other hand, the mean field is derived as, $h_{n}=\int
f(x)\rho_{n}(x)dx$. Relaxation of $\rho_{n}(x)$ can lead the mean
field to a certain critical value, at which the internal bifurcation
occurs in the effective map Eq.(\ref{eq:EffMap}).  For instance, small
difference of the mean field induces one point in the effective map to
be tangential to $y=x$, or one region to be collapsed by crisis.  As a
result, the nature of the invariant measure $\rho_{eq,n}^{(p)}(x)$ of
the effective map changes qualitatively.

With this internal bifurcation, the distribution $\rho_{n}(x)$ is not
actually relaxed to $\rho_{eq,n}^{(p)}(x)$, because 1) the velocity of
change $\rho_{n}(x)$ is finite, and 2) the relaxation of $\rho_{n}(x)$
makes $\rho_{eq,n}^{(p)}(x)$ to be changed qualitatively.
Consequently $\rho_{n}(x)$ oscillates in time.  This is a qualitative
explanation why the mean field does not approach a fixed
point\footnote{The unstable fixed point of the mean field value is
given as $h=\int f(x)\rho_0(x)dx$, where $\rho_0(x)$ is a fixed point
solution of Eq.(\ref{eq:PF}) with $p=1$. The fixed point solution
$\rho_0(x)$ is unstable.\label{foot}} at the thermodynamic limit.

Let us look back to the example in the subsection
\S\ref{ssec:selfconsis} and try to describe the dynamics along the
above scenario.  Population distribution function $\rho_{n}(x)$ at
time step $n$ is plotted with the solid line in Fig.\ref{fig:hist}(and
see the caption in it).  Since the effective nonlinearity parameter
$A$ is near the band merging point of the logistic map, it is useful
to define the two regions as follows.  The effective map given by the
second iterate of map,
\begin{equation}
F_{n}(F_{n-1}(x))=(1-\epsilon)f((1-\epsilon)f(x) +\epsilon
h_{n-1})+\epsilon h_{n},\label{eq:2nd}
\end{equation}
has three unstable fixed points and the middle of these points is
denoted by $x_n^{*}$. $R$ and $L$ denote the region where $x>x_n^{*}$
and $x<x_n^{*}$ respectively. (Based on this definition, we have
calculated the number of elements in the two regions, from which
Fig.\ref{fig:RMSelfCon} in the subsection \S\ref{ssec:selfconsis} is
obtained.)

In this parameter, if the unstable fixed point of the mean field
solution were realized, these two regions collapse due to the
crisis. Hence, these two region are unstable.  (In the bellow
``stable'' or ``unstable'' means that a region, $R$ or $L$, can
attract trajectories or not.) As we have discussed above, however,
dynamics of the mean field modulates the effective map
Eq.(\ref{eq:2nd}), and then, for this parameter regime, there are
three cases: 1) $R$ region is stable and $L$ is unstable, 2) $R$
region is unstable and $L$ is stable, or 3) both $L$ and $R$ regions
are unstable.

The effective map Eq.(\ref{eq:2nd}) can be characterized as invariant
measure of effective map $\rho_{eq,n}^{(2)}(x)$, which depends on the
mean field value. The stability in each region can be seen by the
strength of $\rho_{eq,n}^{(2)}(x)$. The thin lines in
Fig.\ref{fig:hist} are $\rho_{eq,n}^{(2)}(x)$(see the caption in
Fig.\ref{fig:hist}).
In Fig.\ref{fig:hist}(a), $R$ region is unstable and $L$ region is
stable at $n=1$ as is shown by $\rho_{eq,n}^{(2)}(x)$.  Since
$\rho_{n}(x)$ is going to relax to $\rho_{eq,n}^{(2)}(x)$, the
elements in $R$ region moves to $L$ region as is shown in
Fig.\ref{fig:hist}(a) and (b). Indeed the number of elements decreases
in $R$ region and increases in $L$ region in Fig.\ref{fig:SelfCon}(a)
and (b). This change of $\rho_{n}(x)$ continues until the modulation
of the mean field induces crisis of $L$ region at $n=12$ in
Fig.\ref{fig:hist}(c). 
By this destabilization of $L$ region, elements
in $L$ region move to $R$ region so as to relax the distribution
$\rho_{n}(x)$ to $\rho_{eq,n}^{(2)}(x)$ in
Fig.\ref{fig:hist}(d) and (e),
until the next crisis leads to the structure of Fig.\ref{fig:hist}(a),
giving a flow from $R$ region to $L$.

To sum up, distribution function $\rho_{n}(x)$ changes slowly so as to
relax to $\rho_{eq,n}(x)$ and the distribution $\rho_{n}(x)$ changes
until the modulation of the mean field induces internal bifurcation
structure to change qualitatively.  In this example, qualitative
change in internal bifurcation is due to local crisis. By the change
of stability in two regions, $\rho_{n}(x)$ relaxes to a different
region. With the repetition of this stability change the mean field
oscillates quasi-periodically. This mechanism of the stability change
in each band holds for any period-p band(window) regime where elements
are attracted to and repelled from each band region successively with
the internal bifurcation giving rise to crisis.

\section{Bifurcation of Collective Motion}
\label{sec:bif}

\subsection{Bifurcation of Tongue Structures}
\begin{figure}
{\epsfxsize\textwidth\epsfbox{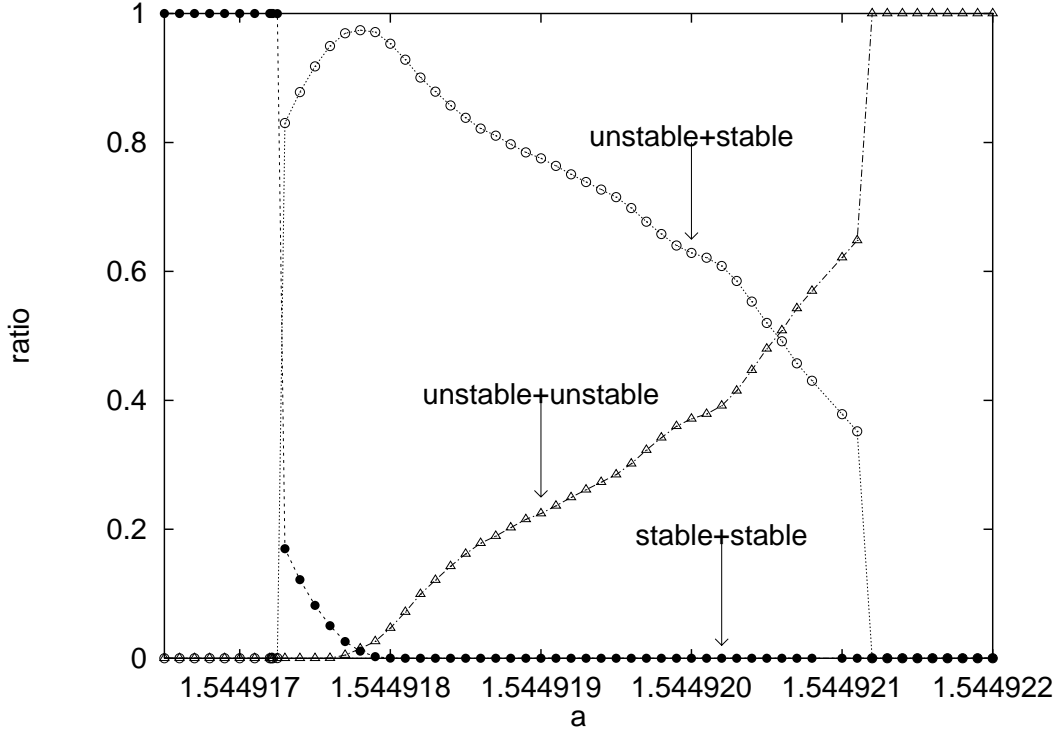}}
\caption{Ratio of the time steps, in which 
the two regions are both stable(``stable+stable''), one of the regions
are stable(``unstable+stable''), and the two regions are both
unstable(``unstable+unstable'').  $\epsilon=0.0005$, $N=10^7$. The
instability is due to the crisis bifurcation in internal bifurcation.}
\label{fig:bifur}
\end{figure}

\begin{figure}
{\epsfxsize\textwidth\epsfbox{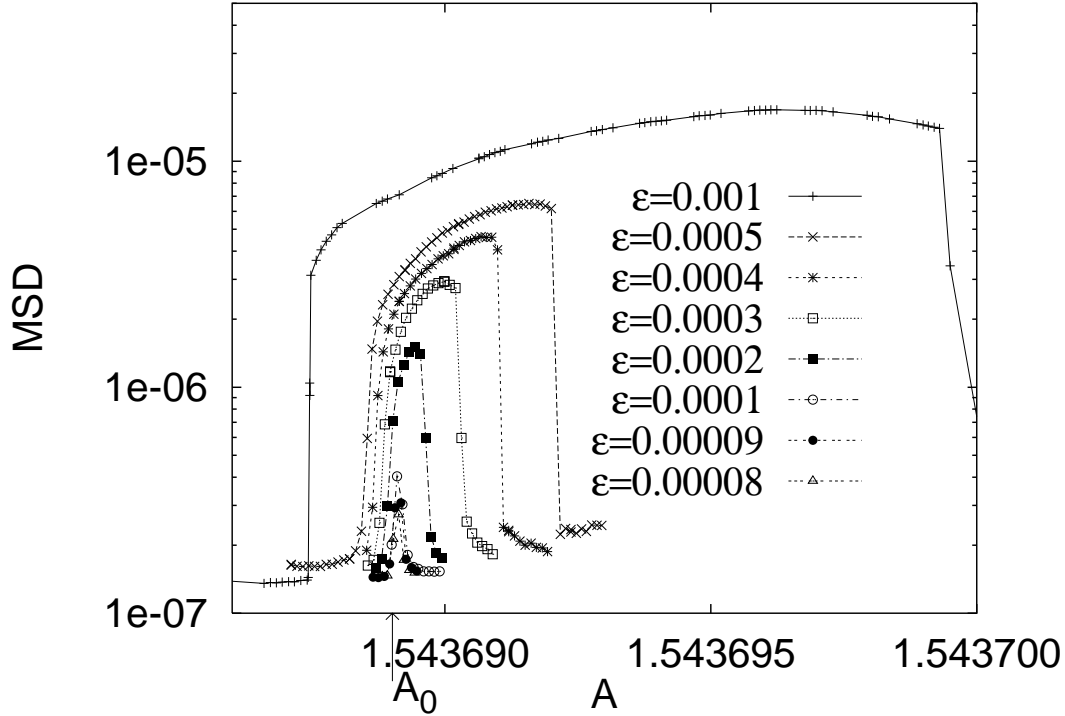}}
\caption{
Mean square deviation (MSD) of the mean field distribution with the
increase of $\epsilon$ plotted as a function of the effective
nonlinearity parameter $A$. The region corresponds to the period-2 tongue
structure.  $A_0$ denotes the band merging point (crisis bifurcation
point of 2-band) of the logistic map ($A_0=1.543689012692076$).}
\label{fig:p2tongue}
\end{figure}

As we have seen in Fig.\ref{fig:hist} in the previous section
(\S\ref{ssec:int}), one of the two regions in the second iterate of
the effective map Eq.(\ref{eq:2nd}) is collapsed due to the crisis at
some time steps and such a region changes in time.  With the increase
of $a$, the time interval of crisis bifurcation becomes longer.  In
Fig.\ref{fig:bifur}, the ratio of the time interval, during which one
of the two regions is unstable and the other is stable, the two
regions are both unstable, and the two regions are both stable, are
plotted with the change of the nonlinearity parameter $a$. For
$a\leq1.5449173$, crisis bifurcation never occurs both in the two
regions, while for $a>1.5449173$, the time interval of crisis is
getting longer. For the parameter beyond $a=1.5449212$, the two
regions is collapsed due to the crisis bifurcation all time steps.
Hence, the period-2 tongue structure starts at the parameter, where
one of the two regions in the second iterate of the effective map
Eq.(\ref{eq:2nd}) is collapsed due to the crisis at some time steps at
$a\cong1.5449173$, and ends up at the parameter where both the two
regions collapse due to the crisis all the time at $a\cong1.5449212$.

Consider an internal bifurcation condition of Eq.(\ref{eq:EffMap})
(for instance, crisis bifurcation or tangent bifurcation in each
element.).  While for $\epsilon=0$ the bifurcation condition holds at
only one point in the nonlinearity parameter, for finite $\epsilon$,
due to the oscillation of the mean field, the internal bifurcation
condition is satisfied for some steps within some interval
$A_{small}(\epsilon)<A(\epsilon)<A_{large}(\epsilon)$ in the parameter
space. Hence, the edge of a tongue structure, corresponding to a
periodic window of logistic map, starts from tangent bifurcation and
crisis bifurcation point at $\epsilon=0$, and each line constitutes
the parameter $A$ and $\epsilon$, where each bifurcation condition
holds at some time steps(Fig.\ref{fig:p2tongue} for period-2 tongue
structure).  Scaling the width of tongue structure will be discussed
lated (in \S\ref{sec:scaling}).

\subsection{Bifurcation in a Tongue Structure}

\begin{figure}
{\epsfxsize\textwidth\epsfbox{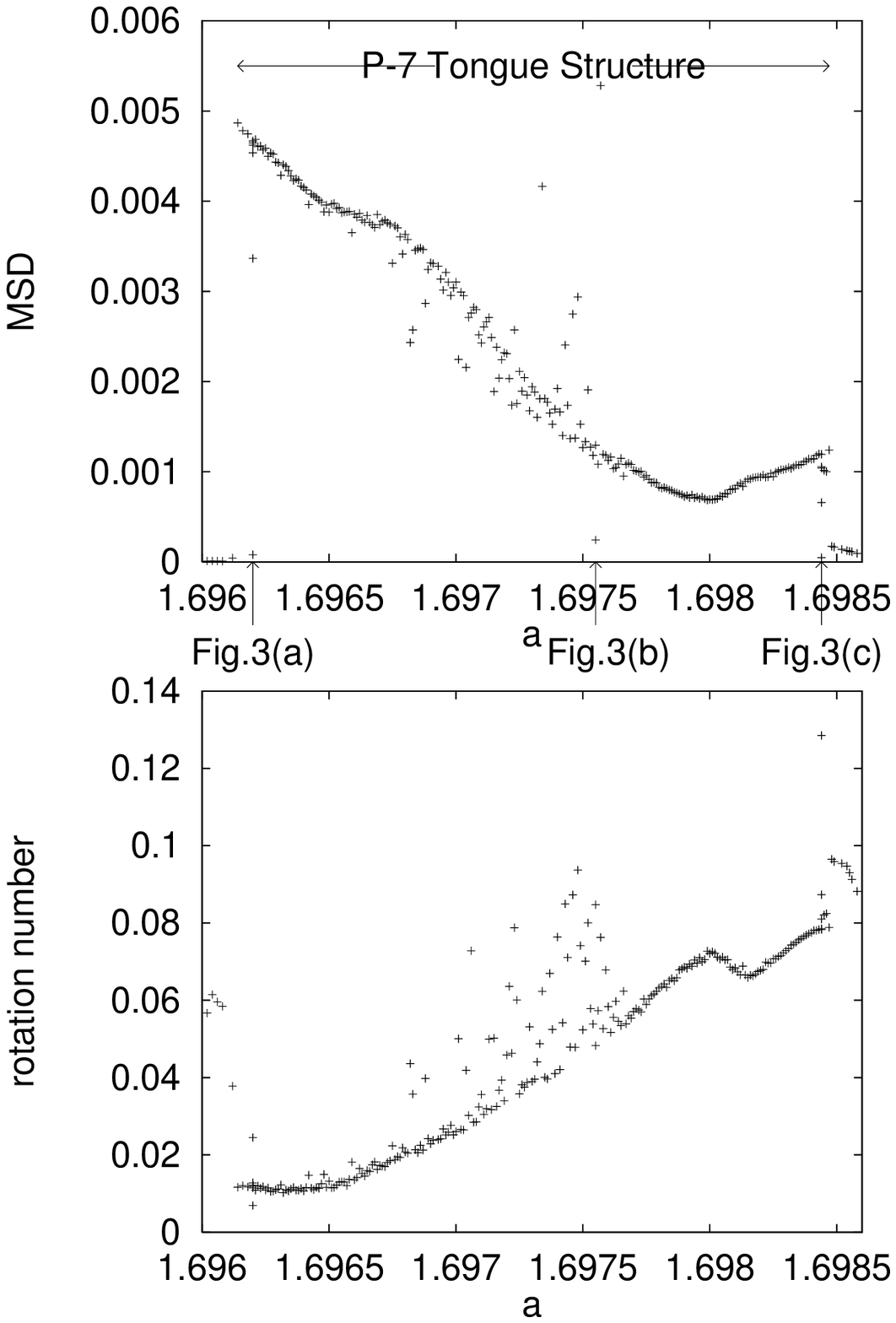}}
\caption{(a)Mean square deviation and (b)rotation number of the mean
field dynamics are plotted as a function of $a$. For the parameter
$a=1.6962$, $1.69755$, and $1.69844$ indicated in the figure (a), see
Fig.\protect\ref{fig:rm.p7} and Fig.\protect\ref{fig:InvAt7_1}}
\label{fig:p7rot}
\end{figure}

\begin{figure}
{\epsfysize.8\textheight\epsfbox{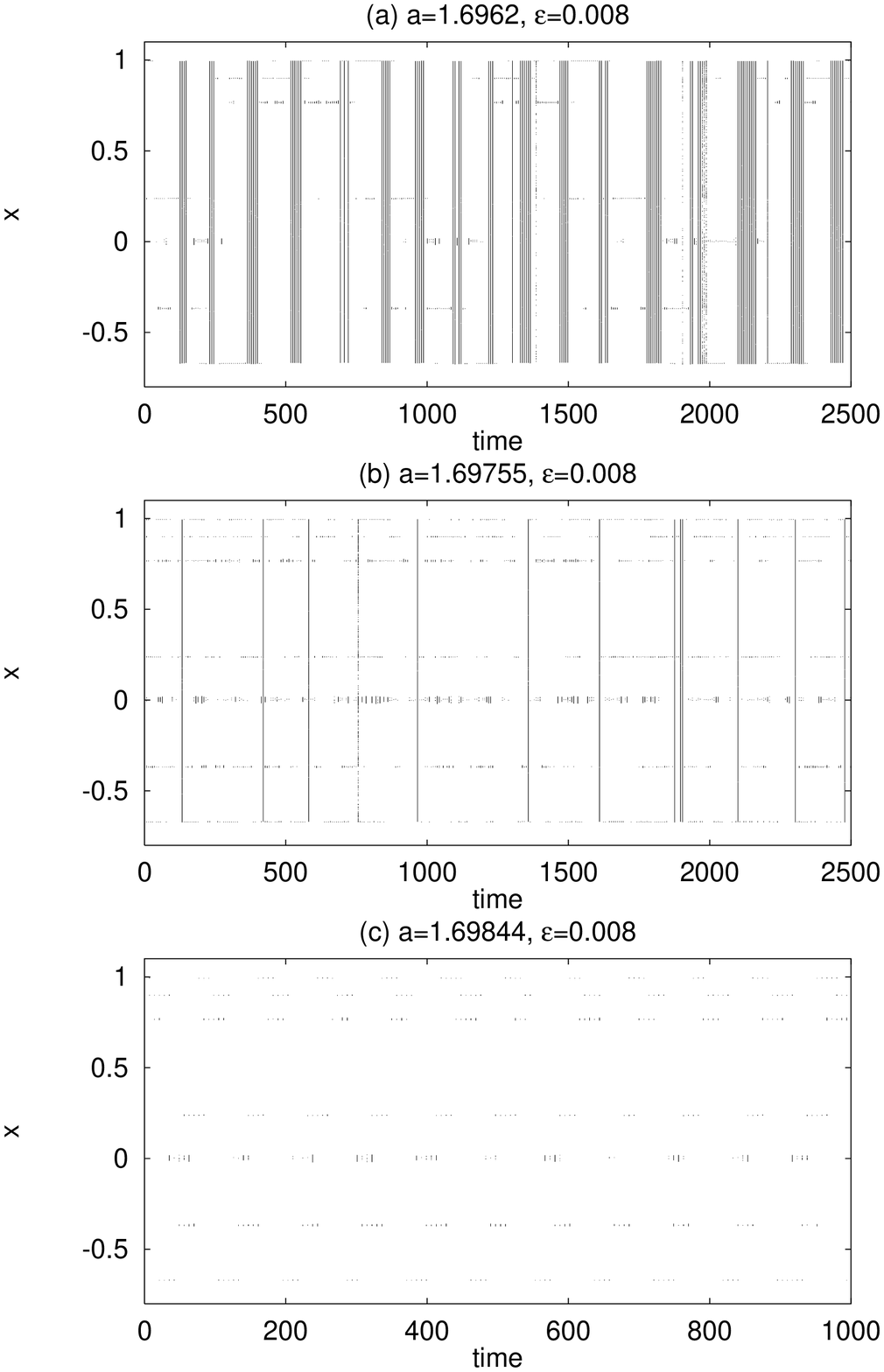}}
\caption{Invariant measure of the effective map
Eq.(\protect\ref{eq:EffMap7}) $\rho_{eq,n}^{(7)}(x)$ is plotted as a
function of time. The horizontal axis is time and the vertical axis is
$x$. In this figure, the region with $\rho_{eq,n}^{(7)}(x)>0$ is
plotted by a solid line. When the whole the region is filled by a
line, none of the seven regions of the map
Eq.(\protect\ref{eq:EffMap7}) cross with $y=x$, and all the regions
are connected as a single attracting set. In (a) and (b) some of the
seven regions of the map Eq.(\protect\ref{eq:EffMap7}) cross with
$y=x$, while the other regions do not.  In (c), on the other hand, all
seven regions of the map Eq.(\protect\ref{eq:EffMap7}) cross with
$y=x$, while some of the seven regions are destabilized by crisis. The
parameters correspond to that of Fig.\protect\ref{fig:rm.p7}.}
\label{fig:InvAt7_1}
\end{figure}

Even within the same tongue structure, we can observe different types
of collective motion.  With the change of the parameter $a$ and
$\epsilon$, the collective dynamics shows a kind of bifurcation.
Since the collective dynamics remains high-dimensional, it is not
described as a standard bifurcation in low-dimensional dynamical
systems.  Here we study a mechanism of such change in the collective
dynamics.

In Section \ref{sec:phenomena}, it is shown that slight increase in
$a$ induces the qualitative change of collective dynamics
(Fig.\ref{fig:rm.p7}). To see this quantitatively, it may be
convenient to measure the rotation number of collective dynamics. In
Fig.\ref{fig:p7rot}, the rotation number is plotted as a function of
$a$.  In the regime plotted in the figure (i.e., between
$a\in[1.69614,1.69847]$ with $\epsilon=0.008$), period-seven tongue
structure is observed. Roughly speaking this period-seven tongue
region is divided into 3 regimes in Fig.\ref{fig:p7rot},
$a\in[1.69614,1.6975]$, $a\in[1.6975,1.698]$ and
$a\in[1.698,1.69847]$. Typical example for each regime are shown in
Fig.\ref{fig:rm.p7}.

To see the mechanism of the difference of dynamics in these parameter
region, it may be useful to adopt the invariant measure
$\rho_{eq,n}^{(7)}(x)$ of the effective map,
\begin{equation}
F^{(7)}_{n}=F_{n+6}\circ F_{n+5}\circ\cdots\circ F_{n+1}\circ
F_{n},\label{eq:EffMap7}
\end{equation}
as we have already introduced in Section
\ref{sec:selfconsistent}. In Fig.\ref{fig:InvAt7_1},
three examples of $\rho_{eq,n}^{(7)}(x)$ are plotted as a function of
time, corresponding to the three regimes mentioned above. In
Fig.\ref{fig:InvAt7_1}(a), seven distinct regions are stabilized
successively.  For this parameter, the effective nonlinearity
parameter $A$ is close to, but smaller than, the tangent bifurcation
point of the period seven window in the logistic map. Therefore if the
fluctuation of the mean field were ignored, none of the seven regions
would be stabilized because the seventh iterate of the logistic map
Eq.(\ref{eq:EffMap7}) does not cross with $y=x$.  With the mean field
dynamics, on the other hand, the effective map Eq.(\ref{eq:EffMap7})
is modified to cross with $y=x$ at a few regions where
$\rho_{eq,n}^{(7)}(x)>0$(for instance between $n=2000$ and $2100$ in
Fig.\ref{fig:InvAt7_1}(a)).  In Fig.\ref{fig:InvAt7_1}(a), two or
three regions are stabilized. After some duration, these regions come
to be destabilized again by crisis (for instance at $n\approx 2100$ in
Fig.\ref{fig:InvAt7_1}(a)).  After the crisis $\rho_{eq,n}^{(7)}(x)$
spreads over the whole region because none of the seven regions of the
map Eq.(\ref{eq:EffMap7}) cross with $y=x$.  Then stabilized regions
switch to different positions.  This process continues successively.

When the parameter $a$ is increased, the number of regions stabilized
by the tangent bifurcation of the map Eq.(\ref{eq:EffMap7}) is
increased (Fig.\ref{fig:InvAt7_1}(b)). In Fig.\ref{fig:InvAt7_1}(b),
5,6,or 7 regions are stabilized successively.  This corresponds to the
second regime in Fig.\ref{fig:p7rot}.

With the further increase of $a$, all the seven regions of the map
Eq.(\ref{eq:EffMap7}) always cross with $y=x$, while some of these
seven regions are destabilized by crisis(Fig.\ref{fig:InvAt7_1}(c)),
as we have shown in section
\ref{sec:selfconsistent}. With the increase of $a$, time duration of
crisis in each seven region is increased, and all the seven regions
start to be destabilized by crisis at the same time step.  Then this
tongue structure ends up and collective dynamics in the p7 tongue
structure is collapsed (at $a=1.69847$).  Then the amplitude of
mean-filed dynamics is reduced smaller to about $1 \over 10$(see
Fig.\ref{fig:p7rot}).

Although we have explained the bifurcation in the internal tongue
structure for the period-7 case, this kind of bifurcation structure is
common to a band region in any period.  For instance, in
Fig.\ref{fig:MSD.scale}(a) with the period-3 tongue structure
(starting from $A\in[1.75, 1.79032]$ at $\epsilon=0.0$) and in the
period-5 tongue structure(starting from $A\in[1.6244,1.6333]$ at
$\epsilon=0.0$), similar bifurcation structure can be observed, where
the change in the number of coexisting stable regions makes such
bifurcation structure.

\subsection{Scaling of Tongue Structures}
\label{sec:scaling}

\begin{figure}
{\epsfxsize\textwidth\epsfbox{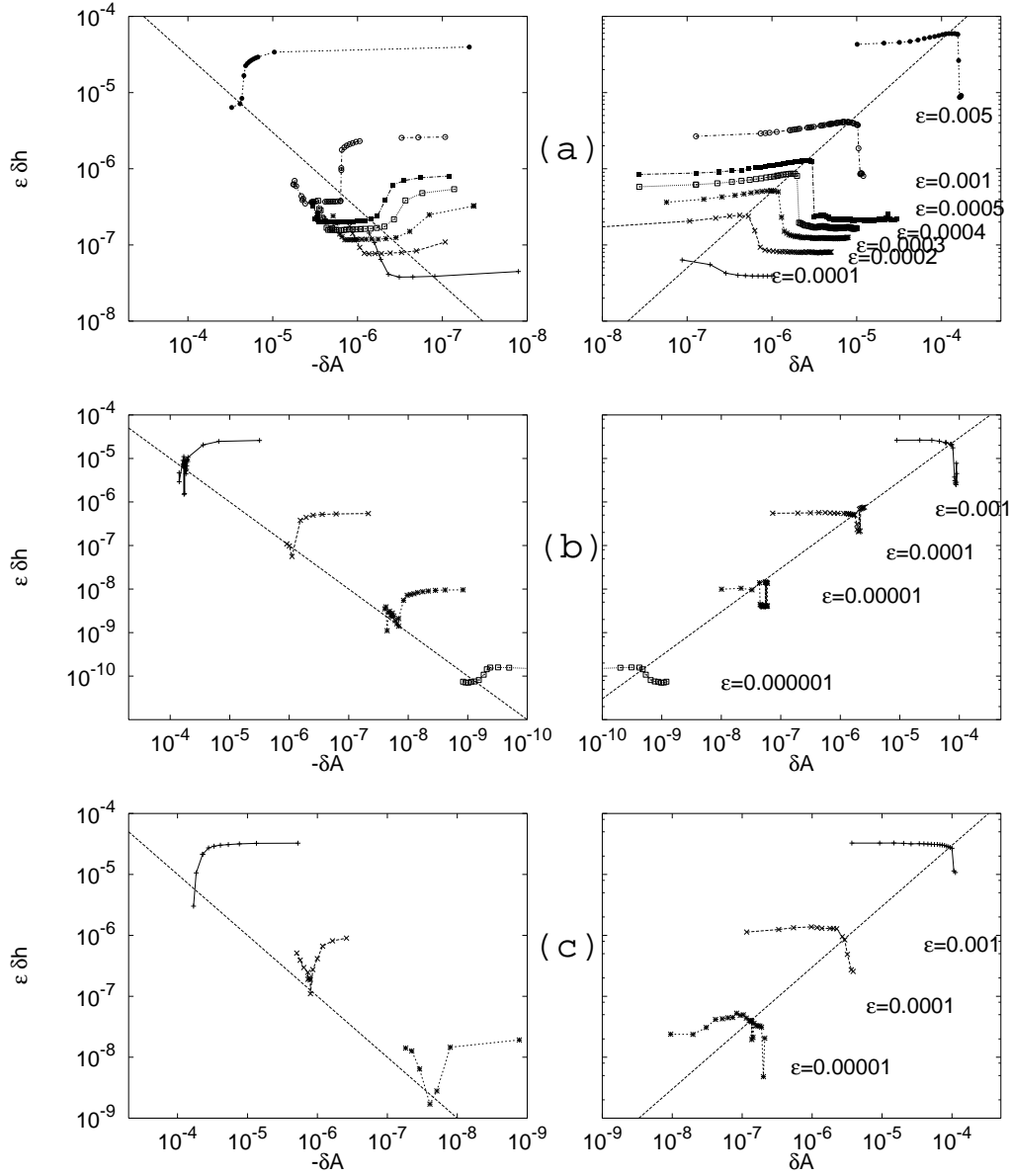}}
\caption{Scaling relation of tongue structure for period-2 (a),
period-3(b), and period-5(c). $\epsilon\delta h$ are plotted as
functions of $\delta A$, where $\delta h=\protect\sqrt{\mbox{MSD}}$
and $\delta A$ indicate the deviation from the band merging
point($A=1.543689012692076$)(Fig.(a)), the crisis bifurcation point of
period-3 window($A=1.790327491999345$)(Fig.(b)), and the crisis
bifurcation point of period-5 window($A=1.633358703691276$)(Fig.(c))
of the logistic map, respectively. Line in each figure is proportional
to $\delta A$. Hence, the edge of $A$ in a tongue structure varies
linearly with $\epsilon\delta h$. The width of a tongue structure
increases proportional to $\epsilon\delta h$.}
\label{fig:scaling}
\end{figure}

\begin{figure}
{\epsfxsize.75\textwidth\epsfbox{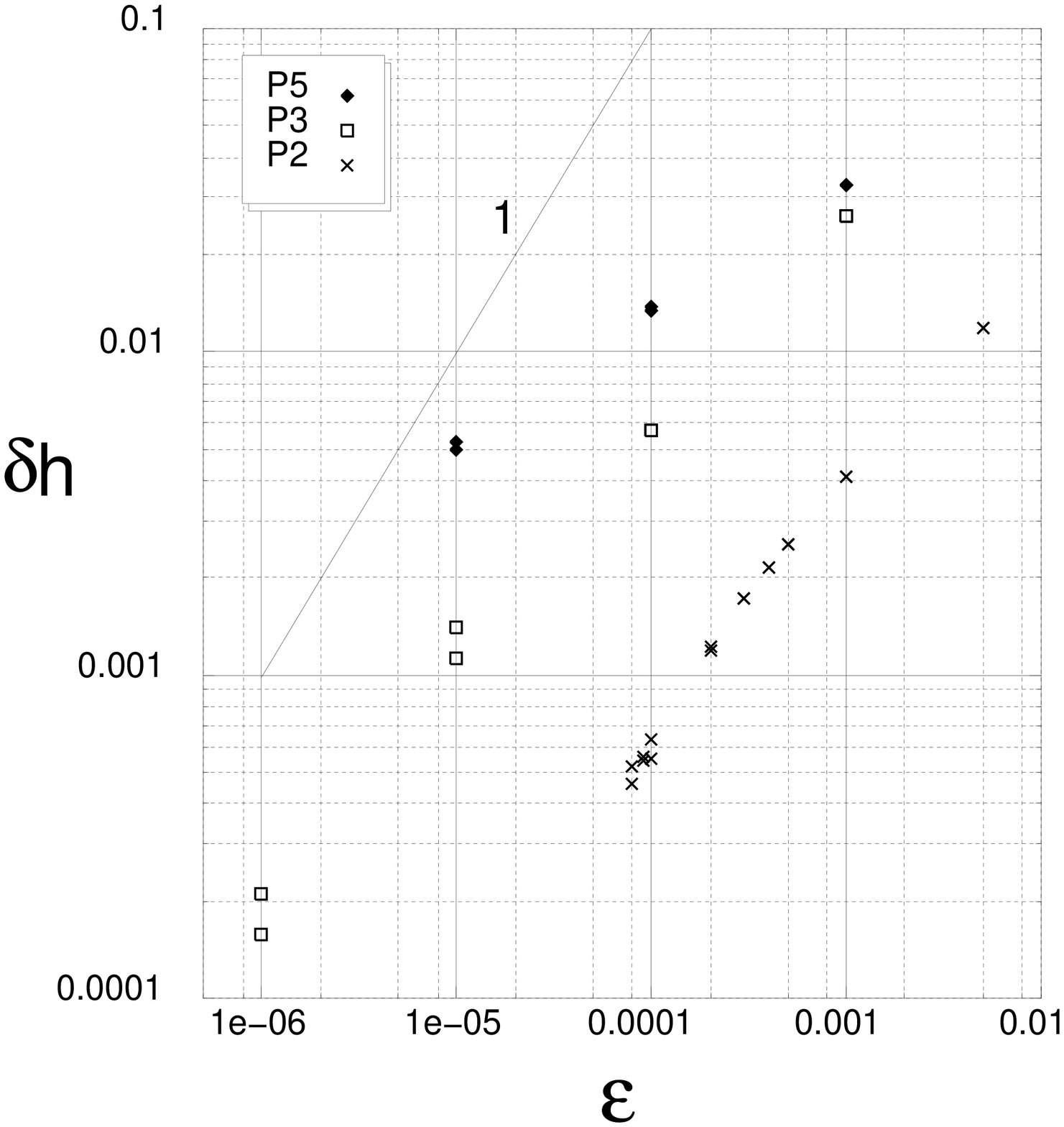}}
\caption{The maximum and next maximum value of $\delta
h=\protect\sqrt{\mbox{MSD}}$ in each tongue structure are plotted.
They are obtained by sampling the data of MSD by changing $A$ in each
tongue with for each $\epsilon$.}
\label{fig:dhvse}
\end{figure}

As we have shown in \S\ref{ssec:Effective Nonlinearity}, the width in
the parameter $A$ of each tongue increases with $\epsilon$ (see
Fig.\ref{fig:MSD.scale}(a)). Here, we discuss the scaling of each
tongue structure.

In general, the effective map for each element is given by
\begin{equation}
F^{(p)}_{n}=F_{n+p-1}\circ F_{n+p-2}\circ\cdots\circ F_{n+1}\circ
F_{n},
\label{eq:effmap2}
\end{equation}
with $F_{n}(x)= (1-\epsilon)(1-a x^2) + \epsilon h_{n}$. By adopting
the effective nonlinearity parameter $A$ and rescaling of $x_n(i)$,
which we have introduced in \S\ref{ssec:Effective Nonlinearity},
$F_{n}(x)$ may be transformed into
\begin{equation}
F_{n}(x)= 1-Ax^2+{\epsilon\cdot\delta h_n \over
1-\epsilon+\epsilon\langle h\rangle},
\end{equation}
where $\delta h_n=h_n-\langle h\rangle$.  As we have mentioned adobe,
an internal bifurcation condition, e.g. crisis bifurcation, and
tangent bifurcation, is satisfied for some steps within some parameter
interval
$A_{small}(\epsilon)<A(\epsilon)<A_{large}(\epsilon)$. (Indeed this
region corresponds to each tongue structure).  Roughly speaking,
$\delta h_n$ varies in time within
$[-\sqrt{\mbox{MSD}}:\sqrt{\mbox{MSD}}]$ for a given parameter.  Thus,
the minimum and maximum parameter of $A$ in a tongue structure is a
function of $-\epsilon\sqrt{\mbox{MSD}}$ and
$+\epsilon\sqrt{\mbox{MSD}}$ respectively, i.e., $A_{small}$ and
$A_{large}$ are given as
\begin{eqnarray}\nonumber
A_{small}(\epsilon)=A_0+A_{1}(-\epsilon\sqrt{\mbox{MSD}}) +
O\left((\epsilon\cdot\sqrt{\mbox{MSD}})^2\right),\\
A_{large}(\epsilon)=A_0+A_{1}(+\epsilon\sqrt{\mbox{MSD}}) +
O\left((\epsilon\cdot\sqrt{\mbox{MSD}})^2\right).
\label{eq:scaling}
\end{eqnarray}
where $A_0$ is a bifurcation parameter of the logistic map, and
$A_{1}$ is positive constant.

For instance, the band merging point of logistic map separates into
two lines as is shown in Fig.\ref{fig:p2tongue}.  In
Fig.\ref{fig:scaling}, tongue structures around the parameter $A$ at
the crisis bifurcation point of the period-2 window(band merging
point), period-3 window, and period-5 window are shown. The linear
scaling relation with $\epsilon\sqrt{\mbox{MSD}}$ is clearly seen as
to the change of $A_{small}$ and $A_{large}$. Thus the width of the
tongue structure grows linearly with $\epsilon\cdot\delta h$.

To obtain the scaling of the tongue structure as a function of
$\epsilon$, we have to know the dependence of MSD on $\epsilon$.  In
Fig.\ref{fig:dhvse}, the growth of square root of the MSD of the mean
field dynamics in a tongue structure with the coupling strength
$\epsilon$ is shown for several tongue structures.  While it has been
pointed out that the amplitude of the mean field dynamics may grow
linearly with $\epsilon$ for globally coupled logistic
map\cite{Kaneko1992,Ershov1997}, Fig.\ref{fig:dhvse} indicates a
possibility that there is a deviation from the linear scaling with
$\epsilon$ for the amplitude of the mean field.  It should be noted
that while we have payed attention mainly to tongue structures
relevant to windows of the logistic map, in the
reference\cite{Ershov1997} such window structures in the logistic map
are out of consideration. In other words, they have focused on the
collective dynamics arising from completely chaotic dynamics in the
logistic map. Note also that our analysis is based on the rescaled
nonlinearity parameter $A$, while the studies in the references
\cite{Kaneko1992,Ershov1997} are based on $a$.  Possible distinction
between the collective motions originated in chaos and window will be
discussed in \S\ref{sec:end} again.

\section{Hysteresis, Multiple Attractors, 
and Coexistence of Different Types of Motion}
\label{sec:hysteresis}

\begin{figure}
{\epsfxsize\textwidth\epsfbox{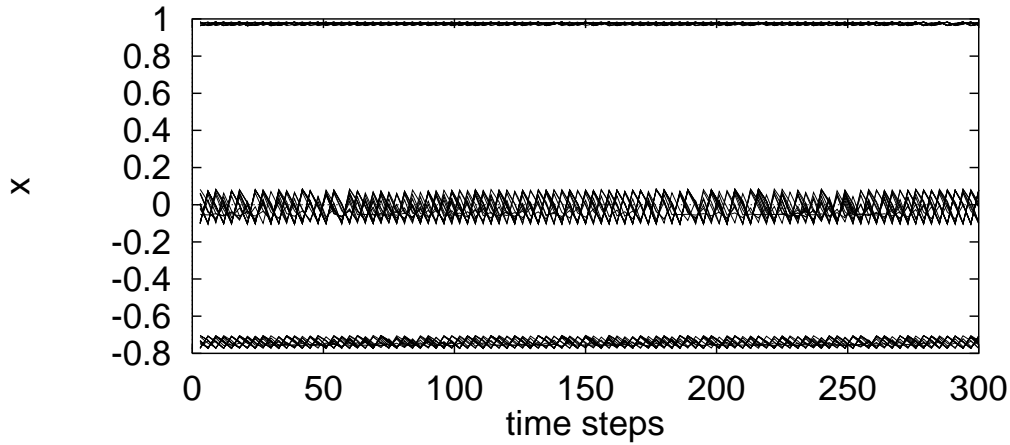}}
\caption{Time series of elements.
Elements are accumulated to three bands.  In the figure time series of
100 elements out of $10^5$ are plotted at every three steps. A lot of
attractors are realized depending on the population ratio to each
band, while elements are desynchronized each other.  The parameters
are $a=1.85$, $\epsilon=0.018$, $N=10^5$.  }
\label{fig:p3.band}
\end{figure}

\begin{figure}
{\epsfxsize.65\textwidth\epsfbox{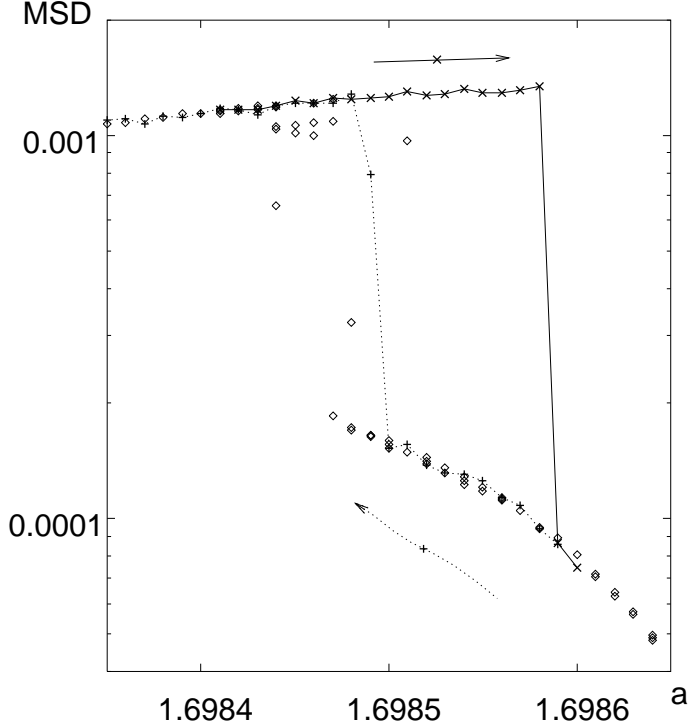}}
\caption{Hysteresis curve observed by increasing or decreasing
progressively the control parameter $a$ while keeping the final state
of a simulation at given $a$ as the initial condition for the
neighboring value $a -\delta a(\times)$ and $a +\delta
a(+)$. $\epsilon=0.008$. The MSD calculated starting from a random
initial condition are also plotted ($\diamond$).}
\label{fig:hysteresis}
\end{figure}

\begin{figure}
{\epsfxsize.9\textwidth\epsfbox{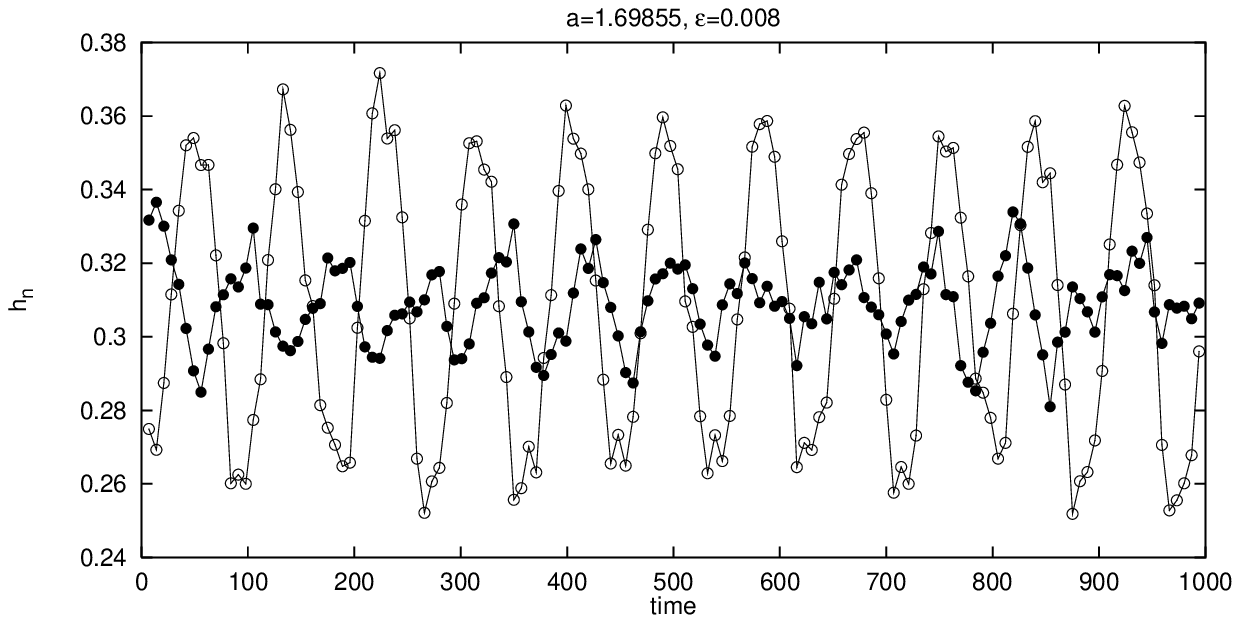}}
{\epsfxsize.45\textwidth\epsfbox{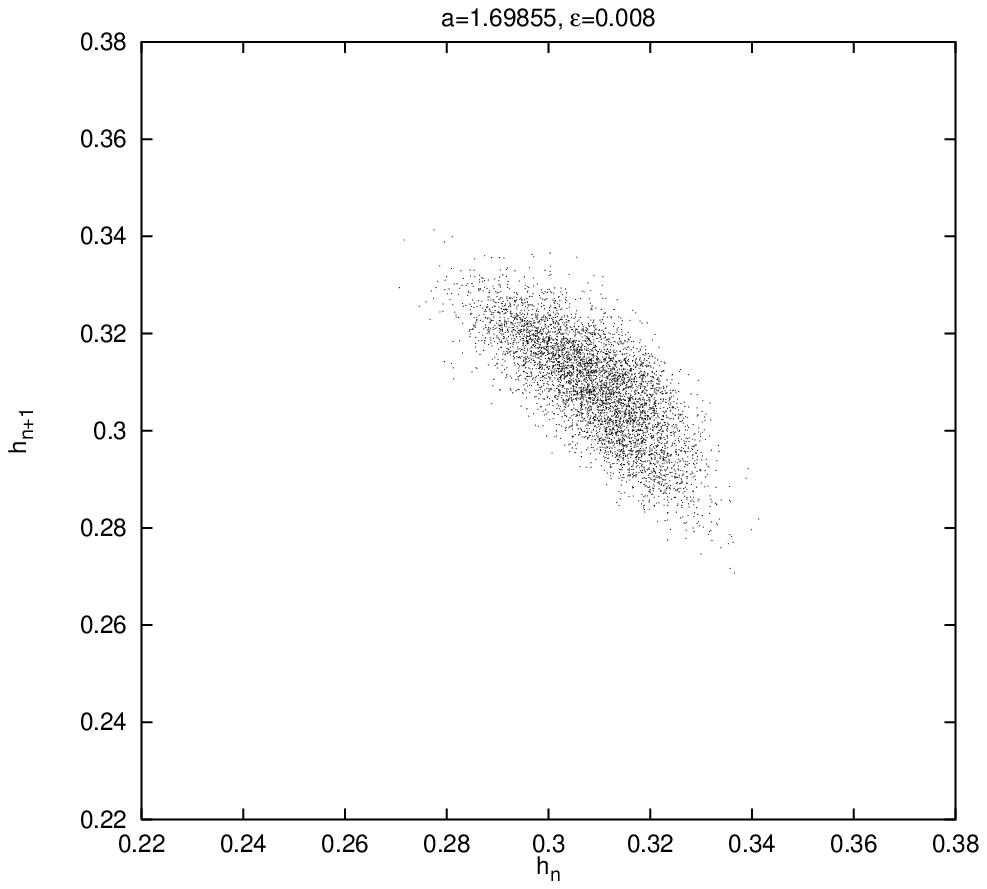}}
{\epsfxsize.45\textwidth\epsfbox{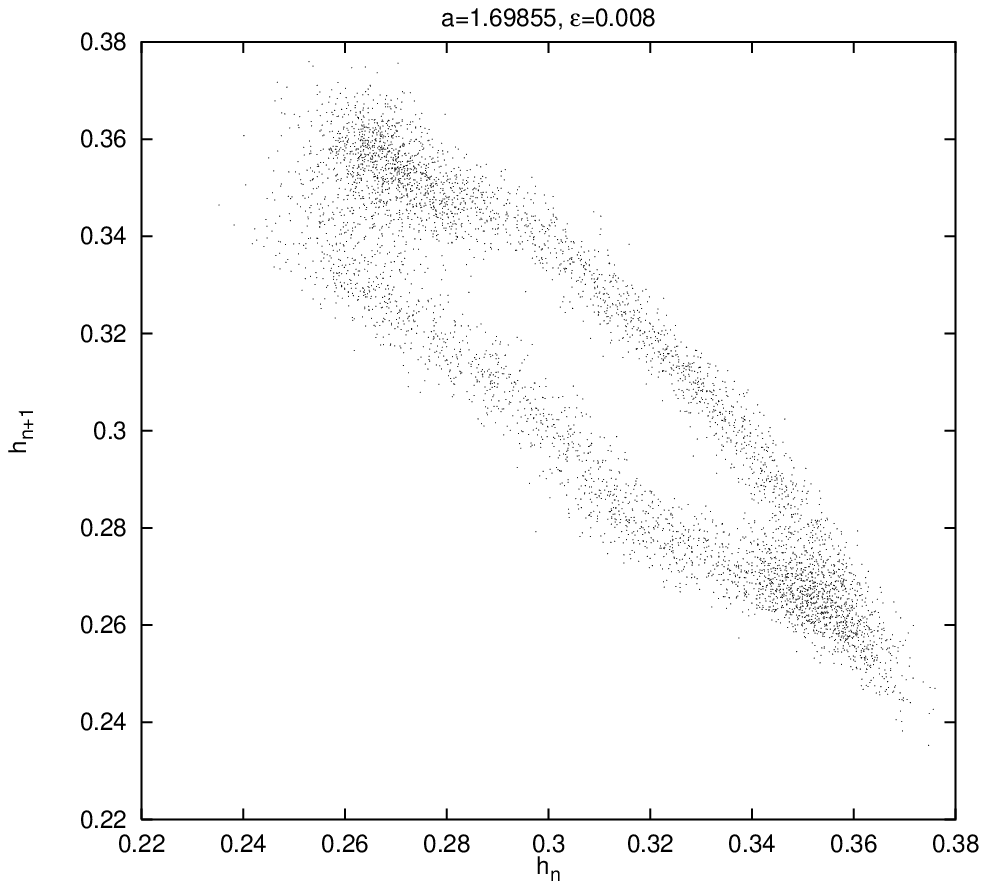}}
\caption{Time series and return map for two attractors in
Fig.\protect\ref{fig:hysteresis}. The parameters are $a=1.69855$,
$\epsilon=0.008$.}
\label{fig:multi}
\end{figure}

\begin{figure}
{\epsfxsize\textwidth\epsfbox{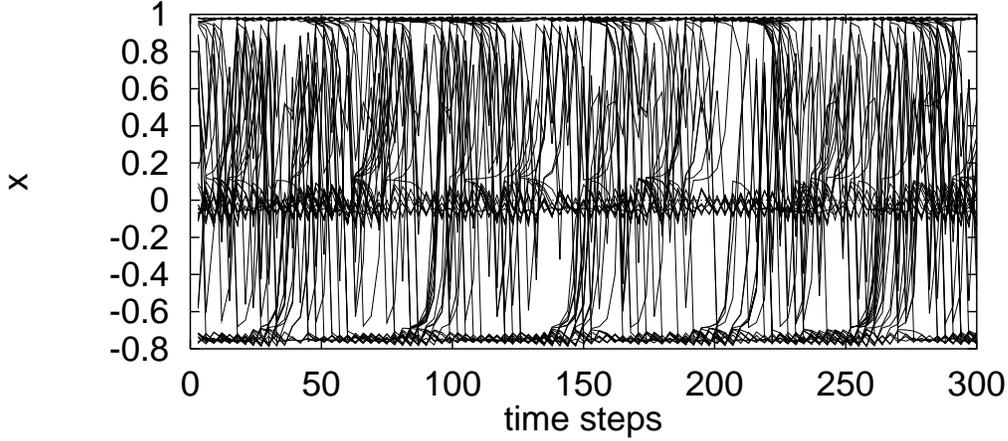}}
\caption{Time series of elements for the same parameter in
Fig.\protect\ref{fig:p3.band} starting from different initial
condition. In contrast with Fig.\protect\ref{fig:p3.band}, elements
spread over $x$. The mean field dynamics for this time series shows
quasi-periodic-like motion. Time series are plotted for 100 elements
at every three steps.  The parameters are $a=1.85$, $\epsilon=0.018$,
$N=10^5$.  }
\label{fig:p3.band2}
\end{figure}

\begin{figure}
{\epsfxsize\textwidth\epsfbox{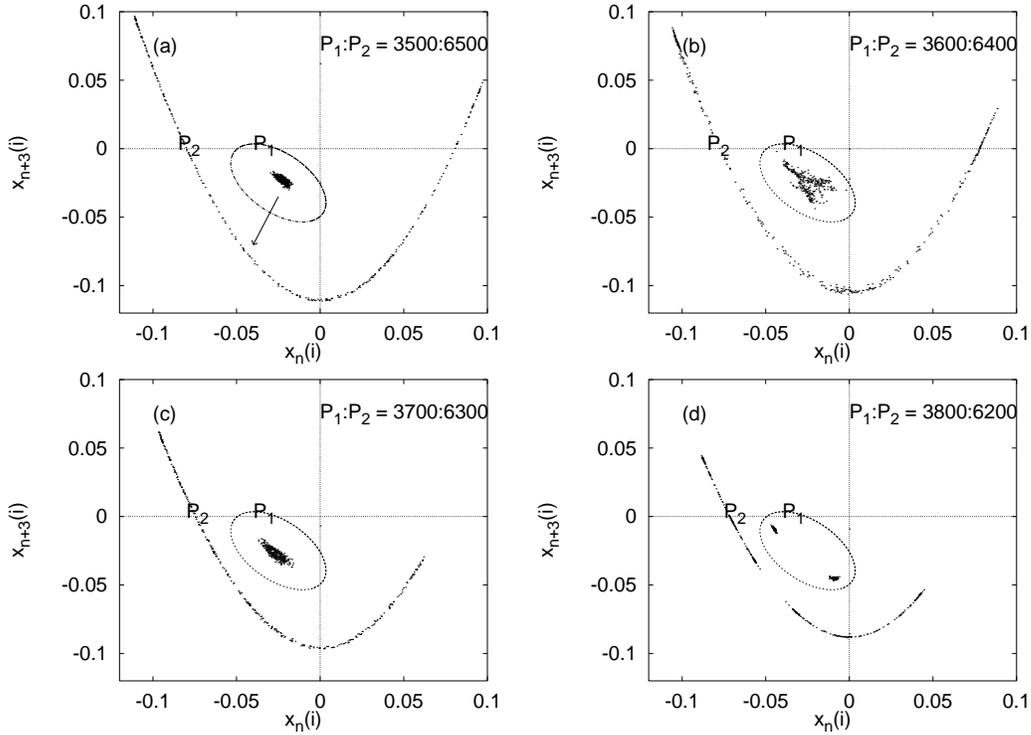}}
\caption{Return maps of element motions to show the coexistence of
different kinds of motions. $\mbox{P}_1$(circled) and $\mbox{P}_2$
represent the element motion belonging to a different band in the
dynamics, respectively. Two kinds of motions (P$_1$, and P$_2$)
coexist.  $\mbox{P}_1:\mbox{P}_2$ in each figure indicates the number
of population in each group.  With the change of ratio, two kinds of
motions are varied.  The parameters are
$a=1.88,\epsilon=0.04,N=10^4$.}
\label{fig:type}
\end{figure}

Even if the control parameters are same, depending on its initial
condition, there can exist more than one attractors of the collective
motion.

Most straightforward examples of multiple attractors are given with
the use of band splitting. At same parameter region, while there is no
mutual synchronization, elements are accumulated to few bands and
never change their band(Fig.\ref{fig:p3.band}).  Therefore, a lot of
attractors are realized depending on the population ratio to each
band, as long as stability conditions are satisfied. For instance, In
Fig.\ref{fig:p3.band}, elements are accumulated into three bands.
(see e.g. \cite{Kaneko1995} for the case with a tent map).

The next example of multiple attractors is concerned with hysteresis
phenomena of collective motion, which can be observed at the edge of
the tongue structure in the parameter space.  In
Fig.\ref{fig:hysteresis}, hysteresis curve of MSD is observed by
increasing or decreasing the control parameter $a$ with the use of the
final state of a simulation at the previous value of $a$ as the next
initial condition. Thus in $a\in[1.69848:1.69858]$, at least two
different attractors of the collective motion coexist. In
Fig.\ref{fig:multi}, time series and return map for each attractor are
shown. Note that, in this case there is no separated bands in contrast
with the previous examples.

The third example of multiple attractors, one attractor has a band
structure (Fig.\ref{fig:p3.band}) and the other not
(Fig.\ref{fig:p3.band2}). For the former attractor, elements are
accumulated in a few bands, while for the latter elements spread over
whole range of $x$.  Moreover for the former type, there exists a lot
of attractors with a different ratio of population in each band, as in
the first example.

Another important topic related to the multiple attractor is the
coexistence of different kinds of element motions.  When the elements
are accumulated into few bands, depending on the ratio of population
in bands, the motion of elements in each band is different. In
Fig.\ref{fig:type}, two kinds of element motions are plotted for
attractors with different population splitting ratio into bands.  At
this parameter, there is a three-band structure, and elements are
accumulated into two groups of these three bands. Note that while
these groups are similar to clusters (cf. \cite{Kaneko1990a}), but the
value of elements $x_{n}(i)$ in each group are different each
other. Depending on the ratio of population in the two groups, two
kinds of element motion coexist.  Relevance of such coexistence to the
problem of cell differentiation is discussed in\cite{furusawa}.

\section{Summary and Discussions}
\label{sec:end}

\begin{figure}
{\epsfxsize.75\textwidth\epsfbox{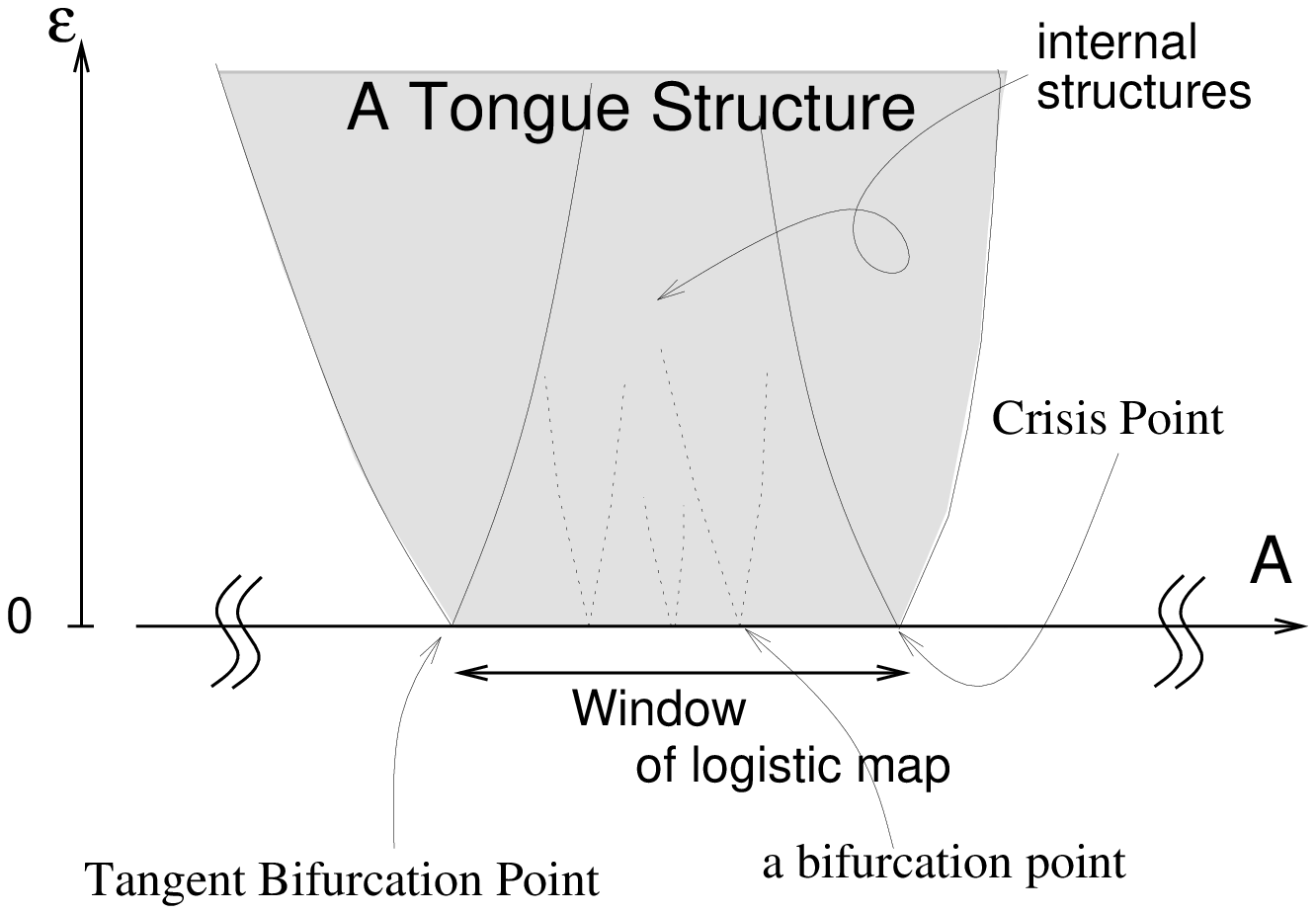}}
\caption{Schematic diagram of tongue bifurcation structure. Regions
that grow from a bifurcation point of the logistic map constitute a
tongue structure. See \S\protect\ref{sec:bif}}
\label{fig:schematic}
\end{figure}

In the present paper, we have studied the collective motion in
desynchronized state of globally coupled logistic maps. It is shown
that the motion with a much longer time scale and lower dimension can
emerge in macroscopic dynamics, such as the mean field dynamics. The
amplitude of collective motion (mean square deviation of the mean
field distribution) is studied by changing the nonlinearity parameter
$a$ and the coupling strength $\epsilon$.  By introducing the
effective nonlinearity parameter $A$ with rescaling of $x_n(i)$,
tongue structure has been detected in ($A,\epsilon$)-plain. Each
tongue structure corresponds to a periodic window of logistic map.

Focusing on the tongue structure, we have demonstrated how such a
collective motion emerges. Self-consistent dynamics between the mean
field dynamics(macroscopic dynamics) and each element(microscopic
dynamics) is found to be formed, so that such a collective motion is
possible. This self-consistent dynamics is formed by the following
circulation: accumulation of elements into some regions leading to the
change in the mean field dynamics, which introduce the stability
change of the the regions, and flow of elements into a different
region, which, again..... This gives internal bifurcation in elements
and in time.

The bifurcation is also seen in the parameter space.  Since the nature
of the internal bifurcation varies with the nonlinearity parameter $a$
in a tongue structure, the number of coexisting regions in $x$
changes, which makes the collective motion qualitatively
different. Hence, in a tongue structure, different kinds of collective
motions have been observed.  (A schematic figure of tongue structure
is presented in Fig.\ref{fig:schematic}).

With the increase of the coupling strength $\epsilon$, each tongue
structure grows in proportion to $\epsilon\cdot\delta h$, where
$\delta h$ is the amplitude of the mean field variation.  Hence the
width of each tongue increases with $\epsilon^2$ if $\delta h
\propto\epsilon$.
In contrast with earlier studies\cite{Kaneko1992,Ershov1997}
supporting this linear scaling, however, our calculation suggests that
the scaling may obey a different power law.

Existence of multiple attractors with a different collective motion
have also been reported. This means that there can be non unique
self-consistent dynamics between microscopic and macroscopic motion.
It should be noted that in the case of the multiple attractors, since
the average mean field is clearly different by attractors, they take
different values of the effective nonlinearity parameter $A$, and are
distinguishable clearly in the ($A,\epsilon$)-plain.

Our tongue structure is based on the windows that exist in the
logistic map.  Since the window exists in any neighborhood in the
parameter space of the logistic map and the width of the tongue
structure increases with the coupling $\epsilon$, the tongue structure
is expected to occupy a relatively large region in the parameter
space.  This is one of the reasons why we have focused our attention
on the collective behavior in the tongue structures.  Still, we have
to note that there is a positive measure in the parameter space of the
logistic map, corresponding to chaos.  Hence, at least at small
coupling regime in our GCM, there are parameters with a positive
measure which do not belong to any tongue structure.  Indeed, we have
observed that the amplitude of the mean-filed variation drops less
than to $1/10$(see Fig.\ref{fig:MSDvsA}), at the parameter where the
tongues structure disappears.  Although no clear structure in the
return map is detected there, this motion again has some hidden
coherence and is distinguishable from noise.  Analytical estimate of
the mean field dynamics by S. V. Ershov, et al.\cite{Ershov1997} is
expected to correspond to such chaos-originated regime.  However, we
need further study to clarify the mechanism of the collective dynamics
there, and characterize the high-dimensional chaos.

While in the desynchronized state elements are completely
desynchronized each other and all the Lyapunov exponents are positive,
for some parameter regime, some kind of predictability may emerge in
the macroscopic variables.  However, since low-dimensional ($O(1)$)
collective dynamics has not been observed, microscopic and macroscopic
dynamics are not separated completely.  This is why we need a
self-consistent description between microscopic variable ($x_n(i)$)
and the mean-field.  On the other hand, it might be also important to
study how such characteristics of the collective motion are reflected
on $N$-dimensional phase space structure, or on microscopic
quantities, such as the Lyapunov spectrum.  With such study, the
mechanism for our collective motion must be clearly distinguishable
from the self-organization mechanism\cite{NicolisPrigogine} or the
slaving principle
\cite{Haken}.  Although we have presented a heuristic way to extract
such self-consistent dynamics in the present paper, it is hoped that a
systematic method to characterize the (high-dimensional) collective
motion will be developed in future\footnote{So far, we have no
conventional tool for detecting the lower dimensional collective
signals out of high dimensional signals. In \cite{Shibata1998}, we
will develop a tool to distinguish and characterize several collective
dynamics in GCM.}.

\section*{Acknowledgment}
Acknowledgment is due to T. Chawanya, S. Morita, and S. Sasa for
valuable discussions of this work. One of the authors (TS) also thanks
E. van Nimweagen and J. P. Crutchfield for fruitful discussions and
valuable comments. A part of the numerical calculation was carried out
at Yukawa Institute Computer Facility.  This work is partially
supported by Grant-in-Aids for Scientific Research from the Ministry
of Education, Science, and Culture of Japan.



\begin{thebibliography}{99}

\bibitem{Watanabe}
S. Watanabe, S. H. Strogatz,
Constants of motion for superconducting Josephson arrays,
{\em Physica D} {\bf 74} (1994) 197-253.;
and references therein.

\bibitem{opt.}
C. Bracikowski, R. Roy, 
Chaos in a multimode solid-state laser system,
{\em Chaos} {\bf 1} (1991) 49; 
F. T. Arecchi, 
Rate processes in nonlinear optical dynamics with many attractors,
{\em Chaos} {\bf 1} (1991) 357.

\bibitem{Aertsen}
Ad Aertsen, M. Erb, G. Plam,
Dynamics of functional coupling in the cerebral cortex:
an attempt at a model-based interpretation,
{\em Physica D} {\bf 75} (1994) 103-128.;
and references therein.

\bibitem{Ko}
E. P. Ko, T. Yomo, I. Urabe,
Dynamics clustering of bacterial population,
{\em Physica D} {\bf 75} (1994) 81-88.;
K. Kaneko, T. Yomo,
Cell division, differentiation and dynamic clustering,
{\em Physica D} {\bf 75} (1994) 89-102.

\bibitem{Godin}
P. J. Godin, T. G. Buchman,
Uncoupling of biological oscillators:
A complementary hypothesis concerning the pathogenesis of multiple
organ dysfunction syndrome,
{\em Crit. Care. Med.} {\bf 24} (1996) 1107-1116.;
T. G. Buchman,
Physiological Stability and Physiologic State,
{\em J. Trauma} {\bf 41} (1996) 599-605.

\bibitem{Kaneko1990a}
K. Kaneko, 
Clustering, Coding, Switching, Hierarchical Ordering, And
Control In A Network Of Chaotic Elements, 
{\em Physica D} {\bf 41} (1990) 137-172.

\bibitem{Kaneko1990b}
K. Kaneko, 
Globally Coupled Chaos Violates the Law of Large Numbers,
{\em Phys. Rev. Lett.} {\bf 65} (1990) 1391.

\bibitem{Kaneko1992}
K. Kaneko, 
Mean field fluctuation of a networks of chaotic elements,
{\em Physica D} {\bf 55} (1992) 368-384.

\bibitem{Perez1992}
G. Perez and H.A. Cerdeira, 
Instabilities and nonstatistical behavior in globally coupled systems,
{\em Phys. Rev. A} {\bf 46} (1992) 7492.

\bibitem{Chate1992}
H. Chat\'e, P. Manneville, 
Collective Behaviors in Spatially Extended Systems with Local
Interactions and Synchronous Updating, 
{\em Prog. Theor. Phys.} {\bf 87} (1992) 1.

\bibitem{Perez1993}
G. Perez, S. Sinha, H. A. Cerdeira, 
Order in the turbulent phase of globally coupled maps,
{\em Physica D} {\bf 63} (1993) 341-349.

\bibitem{Pikovsky1994}
A. S. Pikovsky and J. Kurths, 
Do Globally Coupled Maps Really Violate the Law of Large Numbers?,
{\em Phys. Rev. Lett.} {\bf 72} (1994) 1644;
A.S. Pikovsky and J. Kurths, 
Collective behavior in ensembles of globally coupled maps, 
{\em Physica D} {\bf 76} (1994) 411-419.

\bibitem{Just1995a}
W. Just, 
Bifurcations in Globally Coupled Map Lattices, 
{\em J. Stat. Phys.} {\bf 79} (1995) 429.

\bibitem{Kaneko1995}
K. Kaneko, 
Remarks on the mean field dynamics of networks of chaotic elements,
{\em Physica D} {\bf 86} (1995) 158-170.

\bibitem{Ershov1995}
S. V. Ershov, A. B. Potapov, 
On Mean Field Fluctuations In Globally Coupled Maps,
{\em Physica D} {\bf 86} (1995) 532-558.

\bibitem{Morita1996}
S. Morita, 
Bifurcation in globally coupled chaotic maps, 
{\em Phys. Lett. A} {\bf 211} (1996) 258-264.

\bibitem{Chate1996}
H. Chat\'e, A. Lemaitre, Ph. Marcq, P. Manneville, 
Non-trivial collective behavior in extensively-chaotic dynamical
systems: an update, 
{\em Physica A} {\bf 224} (1996) 447-457.

\bibitem{Shibata1997}
T. Shibata, K. Kaneko, 
Heterogeneity Induced Order in Globally Coupled Chaotic Systems, 
{\em Europhysics Letters} {\bf 38(6)} (1997) 417--422.

\bibitem{Ershov1997} 
S. V. Ershov, A. B. Potapov, 
On Mean Field Fluctuations In Globally Coupled Logistic-Type
Maps. 
{\em Physica D} {\bf 106} (1997) 9-38.

\bibitem{Nakagawa1998}
N. Nakagawa, T. Komatsu, 
Collective motion occurs inevitably in a class of 
populations of globally coupled chaotic elements,
{\em Phys. Rev. E} (1998) in press.

\bibitem{Chawanya}
T. Chawanya, S. Morita, 
On the bifurcation structure of the mean-field fluctuation in the
globally coupled tent map systems. 
{\em Physica D}, in press.

\bibitem{CML}
K. Kaneko, 
Period-doubling of kink-antikink patterns, quasi-periodicity in
antiferro-like structures and spatial intermittency in coupled
logistic lattice --toward a prelude to a ``field theory of chaos''---,
{\em Prog. Theor. Phys}. {\bf 72} (1984) 480; 
Pattern dynamics in spatiotemporal chaos,
{\em Physica D} {\bf 34} (1989) 1; 
Simlulating Physics with Coupled Map Lattice, 
{\it Formation, Dynamics, and Statistics of Patterns}, K. Kawasaki, A. 
Onuki, and M. Suzuki eds. (World Scientific. Singapore, 1990); 
{ \it Theory and Applications of Coupled Map Lattices}, K. Kaneko, ed. 
(Wiley, New York, 1993).

\bibitem{Okuda}
K. Okuda, 
Variety and generality of clustering in globally coupled oscillators,
{\em Physica D} {\bf 63} (1993) 424-436.

\bibitem{Nakagawa1993}
N. Nakagawa and Y. Kuramoto,
Collective chaos in a population of globally coupled oscillators,
{\em Prog. Theor. Phys.} {\bf 89} (1993) 313.

\bibitem{Kaneko1997} 
K. Kaneko, 
Dominance of Milnor Attractors and Noise-induced Selection in a
Multi-attractor System, 
{\em Phys. Rev. Lett,}{\bf 78} (1997) 2736-2739.

\bibitem{CI}
K. Ikeda, K. Matsumoto, and K. Otsuka, 
Maxwell-Bloch turbulence,
{\em Prog. Theor. Phys. Suppl.} 99 (1989) 295; 
I. Tsuda, 
Chaotic itinerancy as a dynamical basis of Hermeneutics in brain and
mind, 
{\em World Futures} {\bf 32} (1992) 313.

\bibitem{Nakagawa1994}
N. Nakagawa and Y. Kuramoto, 
From collective oscillations to collective chaos in a globally coupled
oscillator system, 
{\em Physica D} {\bf 75} (1994) 74-80.

\bibitem{Chabanol}
M. Chabanol, V. Hakim, W. Rappel, 
Collective chaos and noise in the globally coupled complex
Ginzburg-Landau equation, 
{\em Physica D} {\bf 103} (1997) 273-293.

\bibitem{Kuramoto}
Y. Kuramoto,
{\em Surikagaku} {\bf 408} (1997) 5, in Japanese.

\bibitem{Grassberger-Procatia}
P. Grassberger, I. Procatia, 
Characterization of strange attractors, 
{\em Phys. Rev. Lett.}{\bf 50}(1983)346; 
Measuring the strangeness of strange attractors, 
{\em Physica D} {\bf 9} (1983) 189.

\bibitem{furusawa}
C. Furusawa and K. Kaneko,
Emergence of Rules in Cell Society: Differentiation, Hierarchy, and
Stability, 
{\em Bull. Math. Biol.}, in press.

\bibitem{NicolisPrigogine}
G. Nicolis, I. Prigogine,
{\it Self-organization in nonequilibrium systems},
(John Wiley $\&$ Sons, Inc. 1977).

\bibitem{Haken}
H. Haken,
{\it Synergetics},
(Springer-Verlag Berlin Heidelberg, 1976).

\bibitem{Shibata1998}
T. Shibata, K. Kaneko, (1998), in preparation.

\end{thebibliography}
\end{document}